\documentclass[epj,twoside,fleqn,numbook,draft]{svjour}
\usepackage{ifthen}                   % needed for Feyman-Slash macro
\usepackage{exscale}                  % correct scaling of math-symbols
\usepackage[intlimits]{amsmath}       % improved mathematical typesetting
\usepackage{amsfonts}
\usepackage{amssymb,amscd}
\usepackage[dvips]{epsfig}                   
\usepackage{array}
\usepackage{wrapfig}
 
\def\Quadrat#1#2{{\vcenter{\hrule height #2
  \hbox{\vrule width #2 height #1 \kern#1
    \vrule width #2}
  \hrule height #2}}}
\def\dAlemb{\mathop{\kern 1pt\hbox{$\Quadrat{8pt}{0.4pt}$} \kern1pt}}
\def\dAlember{\mathop{\kern 1pt\hbox{$\Quadrat{4pt}{0.4pt}$} \kern1pt}}
\def\Chi{{\mathop{\kern 2pt\vcenter{\hbox{$\chi $}}\kern2pt}}}
\def\dslash{\partial\kern-.5em\slash}
\def\kslash{k\kern-.5em\slash}
\def\slh#1{#1\kern-.5em\slash}
\def\fslash#1{#1 \kern-.5em\slash}

\def\fbar#1{#1\kern-.5em\raise6pt\hbox{\footnotesize /}}

\newcommand{\lsim}{\mathrel{\rlap{\lower4pt\hbox{\hskip1pt$\sim$}}
\raise1pt\hbox{$<$}}}

\newcommand{\Sc}{\scriptstyle}
\newcommand{\be}{\begin{eqnarray}}
\newcommand{\ee}{\end{eqnarray}}

\newcommand{\Pslash}{P\hspace{-.5em}/\hspace{.15em}}
\newcommand{\pslash}{p\hspace{-.5em}/\hspace{.15em}}

\newcommand{\fourint}[1]{\int\!\frac{d^4 #1}{(2\pi)^4}}

\newcommand{\dpartial}[1]{\frac{\partial}{\partial #1}}

\begin{document}

\psfull

{\small
%.... PREPRINT NUMBER ...............................................%
\headnote{
    \hspace*{\fill}{\small\sf UNITUE-THEP-00/08, 
                    \hskip .5cm FAU-TP3-00/07} \\[-6pt]
    \hspace*{\fill}{\small\sf http://xxx.lanl.gov/abs/nucl-th/0006082}
                   }

%...... Title, Authors ..............................................%
\title{\\
 Nucleon Properties in the Covariant\\
              Quark-Diquark Model
%         \thanks{Supported by COSY under contract 41376610
%                 and DFG under contract We1254/4-2}
    }

\author{M.~Oettel\inst{1} \and R.~Alkofer\inst{1} \and
  L.~von~Smekal\inst{2} }

\institute{Universit\"at T\"ubingen, Institut f\"ur Theoretische Physik,
           Auf der Morgenstelle 14, 72076 T\"ubingen, Germany
            \and
           Universit\"at Erlangen-N\"urnberg,
           Institut f\"ur Theoretische Physik III,
           Staudtstr.~7, 91058 Erlangen, Germany}

\date{(30 June 2000)}
\def\makeheadbox{}

\abstract{
In the covariant quark-diquark model the effective Bethe-Salpeter (BS)
equations for the nucleon and the $\Delta$ are solved including scalar {\em
and axialvector} diquark correlations. Their quark substructure is
effectively taken into account in both, the interaction kernel of the BS
equations and the currents employed to calculate nucleon observables.   
Electromagnetic current conservation is maintained. 
The electric form factors of proton and neutron match the
data. Their magnetic moments improve considerably by including axialvector
diquarks and photon induced scalar-axialvector transitions.
The isoscalar magnetic moment can be reproduced,  
the isovector contribution is about 15\% too small. 
The ratio $\mu \,G_E/G_M$ and the axial and strong couplings $g_A$, $g_{\pi
NN}$, provide an upper bound on the relative importance of axialvector
diquarks confirming that scalar diquarks nevertheless describe the
dominant 2-quark correlations inside nucleons. 
}
\maketitle

\pagenumbering{arabic}

\section{Introduction}

High-precision data on nucleon properties in the medium-energy 
range are available by now or will be in near future.
This is especially the case for their electromagnetic form factors.  
From a theoretical point of view the behavior of the form factors
indicates the necessity of a relativistic description of the
nucleon. Non-relativistic constituent models generally fail beyond a momentum
transfer of a few hundred MeV.

From relativistic quantum mechanics of three constituent quarks 
models employing effective Hamiltonian descriptions were deduced. 
However, the necessity for the effective Hamiltonian to comply with the
Poincar\'e algebra and, at the same time, with the covariance properties of
the wave functions leads to fairly complicated constraints 
to ensure the covariance of current operators \cite{Leu78}. 
While some phenomenological studies relax these constraints and 
allow for covariance violations of one-body currents \cite{Szc95},
others put emphasis on the consistent transformation 
of those components of one-body currents that are 
relevant in light-front Hamiltonian dynamics \cite{Chu91,Kap95,Sal95}.
The inclusion of two-body currents was considered within 
a semi-relativistic chiral quark model in the study of 
ref. \cite{Bof99}. Generally, in order to describe 
the phenomenological dipole shape of the electric form factor of the proton,
additional form factors for the constituent quarks need to be introduced in a
phenomenological way in these quantum mechanical models \cite{Sal95,Bof99}.

Models based on the quantum field theoretic bound
state equations for baryons on the other hand have proved capable of
describing the electric form factors of both nucleons quite successfully
without such additional assumptions \cite{Hel97,Blo99,Oet99}. 
Parameterizations of covariant Faddeev amplitudes
of the nucleons for calculating various form factors were explored
in refs.~\cite{Blo99} employing impulse approximate currents, the
field theoretic analog of using one-body currents. 
While the covariance of the corresponding nucleon amplitudes is of course
manifest in the quantum field theoretic models, 
current conservation requires one to go beyond the impulse approximation,
however, also in these studies \cite{Oet99,Bla99,Ish00}.
Furthermore, the invariance under (4-dimensional) translations ramifies into
certain properties of the baryonic bound state amplitudes which are generally
not reflected by the parameterizations but result only for solutions to their
quantum field theoretic bound state equations such as those obtained in
ref. \cite{Oet99}.

The axial structure of the nucleon is known far less precisely than
the electromagnetic one. The theoretical studies have mainly been
focused on the soft point limit. Even though precise experimental data on the
pion-nucleon and the axial form factor 
for finite $Q^2$ are difficult to obtain, and thus practically unavailable, 
it would be very helpful to compare the various theoretical
results (see, {\em e.g.}, refs.\ \cite{Hel97} and \cite{Blo99})
to such data.

In this paper, we investigate the structure of
the nucleon within the covariant quark-diquark model.  
In previous applications of this model, including calculations of  
quark distribution functions \cite{Kus97}
and of various nucleon form factors in the 
impulse approximation \cite{Hel97}, pointlike scalar diquark correlations
were employed.
The octet and decuplet baryon spectrum is described
in ref.\ \cite{Oet98} maintaining pointlike scalar {\em and} axialvector
diquarks. 
A correct description of the nucleons' electric form factors
and radii was obtained in ref.\ \cite{Oet99} by introducing diquark
substructure. We extend this latter calculation   
to non-pointlike axialvector diquarks which allows us to
include the $\Delta$-resonance in the calculations.
The model is fully relativistic, reflects gauge invariance
in the presence of an electromagnetic field, 
and it allows a direct comparison with the semi-relativistic treatment
obtained from the Salpeter approximation (to the relativistic Bethe-Salpeter
equation employed herein). 
This comparison was done in
ref. \cite{Oet00} showing that  both, the  static observables
(magnetic moments, pion-nucleon coupling constant and axial coupling
constant) and the behaviour of the neutron electric form factor 
(or its charge radius, which is particularly sensitive to the specific 
assumptions of any nucleon model), in such a
treatment deviate considerably from the fully relativistic one.
These results indicate that further studies of the 
quantum field theoretic bound state equations for nucleons are worth
pursuing.

In the next section, we briefly review the basic notions
of the quark-diquark model. In Section 3 the model expressions for
the electromagnetic, the pion-nucleon 
and the weak-axial form factors
are derived. 
Starting from the quark vertices and their respective currents
we construct effective diquark vertices and fix their strengths
by resolving the quark-loop structure of the diquarks.
We employ two parameter sets in our calculations. 
One set is chosen to fit nucleon properties alone
whereas the other one includes the mass of the delta resonance.  
While the electric form factors 
are essentially identical 
for both sets, differences occur in the magnetic form factors.
We present our results and discuss their implications in section 4.   
The electromagnetic form factors are thereby compared to experimental data.
In conclusion we comment on future perspectives within this framework.

\section{The Quark-Diquark Model}

Nucleon and delta are modelled as bound states of three constituent quarks.
In order to make the relativistic three-body problem tractable, 
we neglect any 3-particle irreducible interactions between the quarks 
and assume separable correlations in the two-quark channel.
The latter assumption introduces non-pointlike  
diquark correlations. The first assumption
allows to derive a relativistic Faddeev equation for the 6-point quark function
and the assumed separability reduces it to an effective 
quark-diquark Bethe-Salpeter (BS)
equation. In the following, we work in Euclidean space.

In this article we restrict ourselves to
scalar and axialvector diquarks which are introduced, as stated above, via 
the separability assumption for the connected and truncated
4-point quark function:
\begin{eqnarray} \label{Gsep}
G_{\alpha\gamma , \beta\delta}^{\hbox{\tiny sep}}(p,q,P) \,& :=&
 \chi_{\gamma\alpha}(p) \,D(P)\,\bar \chi_{\beta\delta}(q) \; +\; \nonumber
 \\
  & & \hskip 1cm  
\chi_{\gamma\alpha}^\mu(p) \,D^{\mu\nu}(P)
    \bar  \chi_{\beta\delta}^\nu(q)  \; . 
\end{eqnarray}
$P$ is the total momentum of the incoming and the outgoing quark
pair, $p$ and $q$ are the relative momenta between the quarks
in the two channels. 
The propagators of scalar and axialvector diquark in eq. (\ref{Gsep})
are those of free spin-0 and spin-1 particles,
\begin{eqnarray}
D(P) &=& -\frac{1}{P^2+m_{sc}^2} \; , 
 \label{Ds} \\
D^{\mu\nu}(P) &=& -\frac{1}
   {P^2+m_{ax}^2} \left( \delta^{\mu\nu}+  \frac{P^\mu P^\nu}{m_{ax}^2}
\right)  \; .  
 \label{Da}
\end{eqnarray}
Correspondingly, $\chi_{\alpha\beta}(p)$ and $\chi_{\alpha\beta}^\mu(p)$ are
the respective quark-diquark vertex functions.
Here, we maintain only their dominant Dirac structures 
which are multiplied by an invariant function $P(p^2)$ of
the relative momentum $p$ between the quarks to parameterize
the quark substructure of the diquarks.
The Pauli principle then fixes  
this relative momentum to be antisymmetric in the quark momenta 
$ p_\alpha$ and $p_\beta$ \cite{Oet99}, {\it i.e.},
$p=\frac{1}{2}(p_\alpha-p_\beta)$. 
Besides the structure of the vertex functions in Dirac space, they belong 
to the anti-triplet representation in color space, {\em i.e.} they are
proportional to
$\epsilon_{ABD}$ with color indices $A,B$ for the quarks
and $D$ labelling the color of the diquark.
Furthermore, the
scalar diquark is an antisymmetric flavor singlet represented by
$(\tau_2)_{ab}$, and the
axialvector diquark is a symmetric flavor triplet which can be represented
by $(\tau_2\tau_k)_{ab}$. Here, $a$ and $b$ label the quark flavors
and $k$ the flavor of the axialvector diquark. Thus, with all these indices 
made explicit, the vertex functions 
read\footnote{Symbolically denoting the totality of quark indices by 
the same Greek letters that are used as their Dirac indices should not create
confusion. The particular flavor structures are tied to the Dirac
decomposition of the diquarks, color-$\bar 3$ is fixed.}   
\begin{eqnarray}
 \chi_{\alpha\beta}(p)&=&g_s (\gamma^5 C)_{\alpha\beta}\; P(p) \;\;
    \frac{(\tau_2)_{ab}} {\sqrt{2}}\,\frac{\epsilon_{ABD}}{\sqrt{2}} ,
   \label{dqvertex_s} \\
 \chi_{\alpha\beta}^{\mu}(p)&=&g_a (\gamma^\mu C)_{\alpha\beta}\; P(p) \;\;
    \frac{(\tau_2\tau_k)_{ab}} {\sqrt{2}}\,\frac{\epsilon_{ABD}}{\sqrt{2}} .
   \label{dqvertex_a}
\end{eqnarray}
Here, $C$ denotes the charge conjugation matrix;
$g_a$ and $g_s$ are effective coupling constants
of two quarks to scalar and axialvector diquarks, respectively. 

For the scalar function $P(p)$, we employ a simple dipole
form with an effective width $\lambda$ which has proven very successful in
describing the phenomenological dipole form of the electric form factor of
the proton in ref.~\cite{Oet99},   
\begin{equation}
 P(p)= \left( \frac{\lambda^2}{\lambda^2+p^2} \right)^2.
 \label{Pdef}
\end{equation}
This models the non-pointlike nature of the diquarks. It furthermore
provides for the natural ultraviolet regularity of the
interaction kernel in the nucleon and delta BS equations to be derived below.

We can compute the coupling constants $g_s$ and $g_a$ 
by put\-ting the diquarks on-shell
and evaluating the canonical normalization condition 
by using that the vertex functions $\chi_{\alpha\beta}(p)$ and 
$\chi_{\alpha\beta}^\mu(p)$ can be viewed as diquark 
amplitudes with truncated quark legs:
\begin{eqnarray}
 \frac{1}{4}\fourint{p} \; \bar \chi_{\alpha\beta} \; P^\mu \dpartial{P^\mu} 
   G^{(0-)}_{\alpha\gamma,\beta\delta} \; \chi_{\gamma\delta} & 
   \stackrel{!}{=}& 2 m_{sc}^2, \label{normsc} \\
 \frac{1}{4}\fourint{p} \; \left( \bar \chi^{\nu}_{\alpha\beta} \right)^T \;
   P^\mu \dpartial{P^\mu}  
   G^{(0+)}_{\alpha\gamma,\beta\delta} \; 
   \left( \chi^{\nu}_{\gamma\delta} \right)^T 
    &\stackrel{!}{=}& 
                           6 m_{ax}^2. \label{normax}
\end{eqnarray}
Hereby $G^{(0\pm)}_{\alpha\gamma,\beta\delta}$ is the free propagator
for two quarks, symmetrized (+) for the axialvector diquark and
antisymmetrized ($-$) for the scalar diquark \cite{Oet99}.
Furthermore, 
$\left(\chi^{\nu}\right)^T=\chi^\nu - \hat P^\nu (\hat P^\mu \chi^\mu)$
is the transverse part of the vertex function $\chi^{\nu}$ (note
that the pole contribution of the axialvector diquark is transverse to its
total momentum, and the sum over the three polarization states provides 
an extra factor of 3 on the r.h.s. of eq.\ (\ref{normax}). as compared to
eq.~(\ref{normsc})). 
With normalizations as chosen in eqs.\
(\ref{dqvertex_s},\ref{dqvertex_a}) the traces over 
the color and flavor parts yield no additional factors. 

%\subsection{Nucleon amplitudes}

\newpage
\leftline{\it 2.1 Nucleon Amplitudes}
\smallskip

The nucleon BS amplitudes (or wave functions) can be described by 
an effective multi-spinor characterizing the
scalar and axialvector correlations,
\begin{equation}
 \Psi (p,P) u (P,s) \equiv 
    \begin{pmatrix} \Psi^5 (p,P) \\ \Psi^\mu  (p,P) \end{pmatrix} u(P,s).
\end{equation}
$u(P,s)$ is a positive-energy Dirac spinor (of spin $s$), $p$ and $P$ are the
relative and total momenta of the quark-diquark pair, respectively. 
The vertex functions are defined by truncation of the legs,
\begin{equation}
 \begin{pmatrix} \Phi^5  \\ \Phi^\mu \end{pmatrix} = 
    S^{-1}  \begin{pmatrix} D^{-1} & 0 \\ 0 & (D^{\mu\nu})^{-1} \end{pmatrix} 
 \begin{pmatrix} \Psi^5  \\ \Psi^\nu \end{pmatrix} . 
 \label{amp}
\end{equation}
The diquark propagators $D$ and $D^{\mu\nu}$ are defined in
eqs.\ (\ref{Ds},\ref{Da}) and $S^{-1}$ denotes the inverse 
quark propagator in eq. (\ref{amp}). In the present study 
we employ that of a free  constituent quark with mass $m_q$,  
\begin{equation}
 S^{-1}\!(p)= - i \pslash - m_q  \; .               
                          \label{S}                 
\end{equation}                                      
The coupled system of BS equations for the nucleon amplitudes or their vertex
functions can be written in the following compact form, 
\begin{equation}
  \fourint{p'} G^{-1}(p,p',P) 
  \begin{pmatrix}\Psi^5 \\ \Psi^{\mu'}\end{pmatrix}(p',P) =0 \;, 
  \label{bse_nuc}
\end{equation}
in which $G^{-1}(p,p',P)$ is the inverse of the full quark-diquark 4-point
function. It is the sum of the disconnected part and 
the interaction kernel.

Here, the interaction kernel results from the reduction of 
the Faddeev equation for separable 2-quark correlations.
It describes the exchange of the quark with one of those in the 
diquark which is necessary to implement Pauli's principle in the baryon.
Thus,
\begin{eqnarray}
 G^{-1} (p,p',P) &=&\nonumber \\
    &&\hskip -1.2cm 
    (2\pi)^4 \;\delta^4(p-p')\; S^{-1}(p_q)\;
      \begin{pmatrix} D^{-1}\!(p_d)  & 0 \\ 0 & (D^{\mu'\mu})^{-1}\!(p_d)  
         \end{pmatrix} - \nonumber\\
 & &\mbox{\hskip -1.8cm}  \frac{1}{2}
  \begin{pmatrix} - \chi{\Sc (p_2^2) } \; S^T{\Sc (q) }\; \bar\chi{\Sc
  (p_1^2) } &  
     \sqrt{3}\; \chi^{\mu'}{\Sc (p_2^2) }\; S^T{\Sc (q) }\;\bar\chi {\Sc
  (p_1^2) } \\ 
    \sqrt{3}\;\chi{\Sc (p_2^2) }\; S^T{\Sc (q) }\;\bar\chi^{\mu}{\Sc (p_1^2) }
    &  \chi^{\mu'}{\Sc (p_2^2) }\; S^T{\Sc (q) }\;\bar\chi^{\mu}{\Sc (p_1^2) }
     \end{pmatrix} \; . %\nonumber 
 \label{Kdef}
\end{eqnarray}
Herein, the flavor and color factors have been taken into account explicitly,
and $\chi, \, \chi^{\mu}$ stand for the Dirac structures of the
diquark-quark vertices (multiplied by the invariant function
$P(p_{1,[2]}^2)$, see eq.\ (\ref{Pdef})). 
The freedom to partition the total momentum between quark and diquark
introduces the parameter $\eta \in [0,1]$ with $p_q=\eta P+p$ and
$p_d=(1-\eta)P - p$. The momentum of the exchanged quark is then given by
$q=-p-p'+(1-2\eta)P$. The relative momenta of the quarks in the diquark
vertices  $\chi$ and  $\bar\chi$ are $p_2=p+p'/2-(1-3\eta)P/2$ and 
$p_1=p/2+p'-(1-3\eta)P/2$, respectively.
Invariance under (4-dimensional) translations implies that for
every solution  $\Phi(p,P;\eta_1)$ of the BS equation there exists 
a family of solutions of the form $\Phi(p+(\eta_2-\eta_1)P,P;\eta_2)$.

Using the positive energy projector with
nucleon bound-state mass $M_n$,
\begin{equation}
 \Lambda^+= \frac{1}{2}\left( 1+ \frac{\Pslash}{iM_n}\right),
\end{equation}
the vertex functions can be decomposed into their most general Dirac
structures, 
\begin{eqnarray}
 \Phi^5(p,P)&=& (S_1 +\frac{i}{M_n}\pslash S_2) \Lambda^+, \qquad 
   \label{phi5deco}\\
 \Phi^\mu(p,P)&=& \frac{P^\mu}{iM_n} (A_1 +\frac{i}{M_n}\pslash A_2) \gamma_5
      \Lambda^+  + \nonumber \\
              & & \gamma^\mu (B_1 +\frac{i}{M_n}\pslash B_2) 
                   \gamma_5\Lambda^+ + \label{phimudeco}\\
              & & \frac{p^\mu}{iM_n}( C_1 + \frac{i}{M_n}\pslash C_2) 
                  \gamma_5\Lambda^+ \; \nonumber .
\end{eqnarray}
In the rest frame of the nucleon, $P = ( \vec 0 , iM_n)$,  
the unknown scalar functions $S_i$ and $A_i$ are functions of $p^2=p^\mu
p^\mu$ and of the angle variable $z= \hat P \cdot \hat p$,
the cosine of the (4-dimensional) azimuthal angle of $p^\mu$.
Certain linear combinations of these eight covariant components then 
lead to a full partial wave decomposition, see ref.\ \cite{Oet98} for more
details and for examples of decomposed amplitudes assuming pointlike
diquarks. These nucleon amplitudes have in general a  considerably 
broader extension in momentum space
than those obtained herein with including the quark
substructure of diquarks, however.

The BS solutions are normalized by the canonical condition
\begin{eqnarray}
  M_n \Lambda^+ \;& \stackrel{!}{=}&
  -\int \frac{d^4\,p}{(2\pi)^4}
  \int \frac{d^4\,p'}{(2\pi)^4} \label{normnuc}\\
   & &\bar \Psi(p',P_n) \left[ P^\mu \frac{\partial}{\partial P^\mu}
    G^{-1} (p',p,P) \right]_{P=P_n} \hskip -.5cm  \Psi(p,P_n) \; . 
  \nonumber 
 \end{eqnarray}

%\subsection{ Delta amplitudes}
\leftline{\it 2.2 Delta Amplitudes}
\smallskip

The effective multi-spinor for the delta baryon representing
the BS wave function can be characterized as $\Psi_\Delta^{\mu\nu}(p,P)
u^\nu(P)$ where $u^\nu(P)$ is a Rarita-Schwinger spinor.
The momenta are defined analogously to the nucleon case. 
As the delta state is flavor symmetric, only the axialvector
diquark contributes  and,  accordingly, the corresponding BS equation reads,
\begin{eqnarray}
  \fourint{p'} G^{-1}_\Delta (p,p',P) 
  \Psi_\Delta^{\mu'\nu}(p',P) =0 \; ,
  \label{bse_del}
\end{eqnarray}
where the inverse quark-diquark propagator $G^{-1}_\Delta$ in the
$\Delta$-channel is given by
\begin{eqnarray}
  G^{-1}_\Delta(p,p',P) &=&  (2\pi)^4 \delta^4(p-p')\; S^{-1} (p_q) \;
        (D^{\mu\mu'})^{-1} (p_d) + \nonumber \\ 
    & &  \chi^{\mu'}(p_2^2)\; S^T(q)\;\bar\chi^\mu(p_1^2).
\end{eqnarray}
The general decomposition of the corresponding vertex function
$\Phi^{\mu\nu}_\Delta$, obtained as in eq.\ (\ref{amp})
by truncating the quark and diquark legs of the BS wave function
$\Psi_\Delta^{\mu\nu}$, reads
\begin{eqnarray}
 \Phi^{\mu\nu}_\Delta (p,P) &=& (D_1 + \frac{i}{M_\Delta} \pslash D_2) \
                        \Lambda^{\mu\nu} + \nonumber \\
   & &  \frac{P^\mu}{iM_\Delta} (E_1 + \frac{i}{M_\Delta} \pslash E_2) 
        \frac{p^{\lambda T}}{iM_\Delta} \Lambda^{\lambda\nu} + 
        \label{Deldec}\\
   & &  \gamma^\mu (E_3 + \frac{i}{M_\Delta} \pslash E_4 ) 
        \frac{p^{\lambda T}}{iM_\Delta} \Lambda^{\lambda\nu} + \nonumber \\
   & &  \frac{p^\mu}{iM_\Delta} ( E_5 + \frac{i}{M_\Delta} \pslash E_6) 
        \frac{p^{\lambda T}}{iM_\Delta} \Lambda^{\lambda\nu} \; . \nonumber 
\end{eqnarray} 
Here, $\Lambda^{\mu\nu}$ is the Rarita-Schwinger projector,
\begin{eqnarray}
  \Lambda^{\mu\nu} \!= \Lambda^+
             \left(
             \delta^{\mu\nu}-\frac{1}{3}\gamma^\mu\gamma^\nu
             +\frac{2}{3} \frac{P^\mu P^\nu}{M_\Delta^2} - 
             \frac{i}{3} \frac{P^\mu\gamma^\nu-P^\nu\gamma^\mu}{M_\Delta} 
              \right)
             \end{eqnarray}
 which obeys the
constraints
\begin{equation}
  P^\mu \Lambda^{\mu\nu} = \gamma^\mu \Lambda^{\mu\nu} =0.
\end{equation}
Therefore, the only  non-zero components arise from the 
contraction with the transverse relative momentum
$p^{\mu T}=p^\mu - \hat P^\mu (p\cdot \hat P)$.
The invariant functions $D_i$ and $E_i$ in eq.~(\ref{Deldec}) again depend
on $p^2$ and $\hat p\cdot \hat P$.
The partial wave decomposition in the rest frame is given in
ref.\ \cite{Oet98}, and again, the $\Delta$-amplitudes from pointlike diquarks
of \cite{Oet98} are wider in $p^2$ than those obtained herein.

%\subsection{Solution}
\bigskip
\bigskip
\leftline{\it 2.3 Solutions for BS Amplitudes of Nucleons and $\Delta$}
\smallskip

The nucleon and $\Delta$ BS equations are solved in the baryon rest frame
by expanding the unknown scalar functions 
in terms of Chebyshev polynomials \cite{Oet98,Oet99}. 
Iterating the integral equations yields a certain eigenvalue 
which by readjusting the parameters of the model is tuned to one.
Altogether there are four parameters, the quark mass
$m_q$, the diquark masses $m_{sc}$ and $m_{ax}$ and the diquark
width $\lambda$.

\begin{table}[t]
\begin{center}
\begin{tabular}{rl|ll}
 & & Set I & Set II \\[4pt]
% & & & \\ 
\hline
 $m_q$ & [GeV]& 0.360 & 0.425 \\
 $m_{sc}$ & [GeV]& 0.625 & 0.598 \\
 $m_{ax}$ & [GeV]& 0.684 & 0.831 \\
 $\lambda$ & [GeV]& 0.95 & 0.53 \\ \hline
 $g_s$& & 9.29 & 22.10 \\
 $g_a$& & 6.97 & 6.37\\
 $M_\Delta$ & [GeV]& 1.007 & 1.232 \\
 $M_n$ & [GeV]& 0.939 & 0.939 
\end{tabular}
\caption{The two parameter sets employed in the calculations herein together
with the values of couplings and bound state masses obtained with these sets.}
\label{pars}
\end{center}

\end{table}

In the calculations presented herein we shall illustrate the consequences of
our present model assumptions with two different parameter 
sets as examples which emphasize slightly different aspects.
For Set I, we
employ a constituent quark mass of $m_q=0.36$ GeV which is
close to the values commonly used by non- or semi-relativistic
constituent quark models. Due to the free-particle poles in the
bare quark and diquark propagators used presently in the model, the
axialvector diquark mass is below 0.72 GeV and the delta mass
below 1.08 GeV. On the other hand, nucleon {\em and} delta masses 
are fitted by Set II, {\em i.e.}, the parameter space 
is constrained by these two masses.
In particular, this implies $m_q>0.41$
GeV. Both parameter sets together with the corresponding values resulting 
for the effective diquark couplings and baryon masses are given in
table~\ref{pars}.  

Two differences between the two sets
are important in the following: The strength of the axialvector correlations
within the nucleon is rather weak for Set II, since the scalar diquark
contributes 92\% to the norm integral of eq.~(\ref{normnuc}) while
the axialvector correlations and scalar-axialvector
transition terms together give rise only to the remaining 8\% for this
set. For Set I, the fraction of the scalar correlations 
is reduced to 66\%, the axialvector correlations are therefore expected 
to influence nucleon properties more strongly for Set I than for Set II.
Secondly, the different constituent quark masses affect 
the magnetic moments. We recall that in non-relativistic quark models
the magnetic moment is roughly proportional to $M_n/m_q$ and that 
most of these models thus employ constituent masses around 0.33 GeV.

\section{Observables}

%\subsection{Electromagnetic Form Factors}

\refstepcounter{subsection}
\leftline{\it 3.1 Electromagnetic Form Factors}
\smallskip

\begin{figure}[t]
\begin{center}
 \epsfig{file=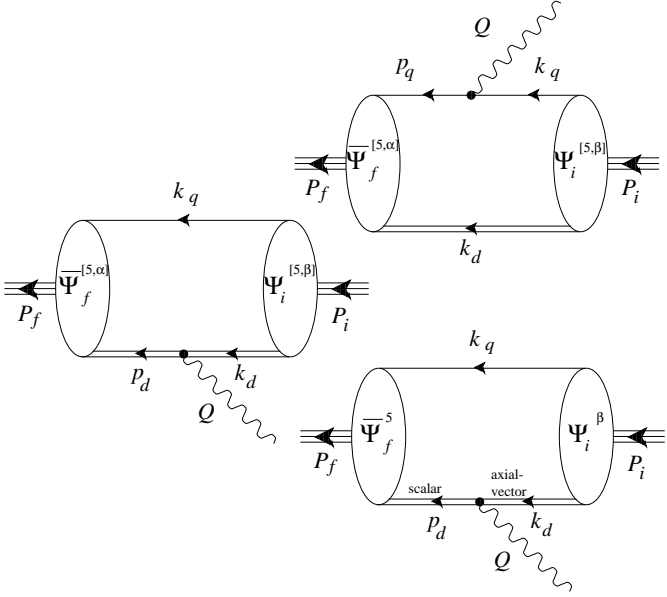,width=\columnwidth}
\end{center}
\caption{Impulse approximate contributions to the electromagnetic
current. For the scalar-axialvector transition, a diagram analogous to
the third one (with initial and final nucleon states interchanged) 
has to be computed.}
\label{impulse}
\end{figure}

The Sachs form factors $G_E$ and $G_M$ can be extracted from the solutions
of the BS equations using the relations
\begin{equation}
 G_E =\frac{M_n}{2P^2} \; \mbox{Tr} \langle J^\mu \rangle P^\mu, \quad
 G_M =\frac{iM_n^2}{Q^2}  \; \mbox{Tr}\langle J^\mu \rangle \gamma^\mu_T\;,
\end{equation}
where $P=(P_i+P_f)/2$, $\gamma^\mu_T=\gamma^\mu- 
\hat P^\mu \hat \Pslash$, and the spin-summed matrix element $ \langle J^\mu
\rangle $ is given by   
\begin{eqnarray}
 \langle J^\mu \rangle &\equiv&  \langle P_f,s_f | J^\mu | P_i,s_i \rangle \,
 \sum_{s_f,s_i} u(P_f,s_f) \bar{u}(P_i,s_i)  \nonumber\\
   &=& \fourint{p_f} \fourint{p_i}  
       \bar \Psi (P_f,p_f) \; J^\mu \; \Psi (P_i,p_i).  \label{curmelt}
\end{eqnarray}
The current $J^\mu$ herein is obtained as in ref.~\cite{Oet99}.
It represents a sum of all possible couplings of the photon to the inverse
quark-diquark propagator $G^{-1}$ given in eq.\ (\ref{Kdef}). This
construction which ensures current conservation can be systematically derived  
from the general ``gauging technique'' employed in refs. \cite{Bla99,Ish00}.

The two contributions to the current that arise from
coupling the photon to the disconnected part of $G^{-1}$, the 
first term in eq.\ (\ref{Kdef}), yield the impulse approximate couplings  
to quark and diquark. They are graphically represented 
by the middle and the upper diagram
in figure~\ref{impulse}. The corresponding kernels, to be 
multiplied by the charge of the respective 
quark or diquark upon insertion into the r.h.s. of eq.~(\ref{curmelt}), read,
\begin{eqnarray}
 J^\mu_q &=& (2\pi)^4\, \delta^4(p_f-p_i-\eta Q) %\, q_q\, 
              \Gamma^\mu_q \,
              \tilde D^{-1} (k_d), \label{iaqv}\\
 J^{\mu}_{sc[ax]} &=& (2\pi)^4\, \delta^4(p_f-p_i+(1-\eta) Q) %\, q_{dq} \,
              \Gamma^{\mu,[\alpha\beta]}_{sc[ax]}\, 
              S^{-1} (k_q). \label{iadqv}
\end{eqnarray} 
Here, the inverse diquark propagator $\tilde D^{-1}$ comprises
both, scalar and axialvector components.
The vertices in eqs.~(\ref{iaqv}) and
(\ref{iadqv}) are the ones for a free quark, a spin-0 and a spin-1 particle,
respectively,
\begin{eqnarray}
 \Gamma^\mu_q \!&=&\! -i \gamma^\mu, \qquad  \Gamma^{\mu}_{sc} \; =\;
 -(p_d+k_d)^\mu, \qquad \mbox{and}  \label{qandsdqv}\\
 \Gamma^{\mu,\alpha\beta}_{ax} \!&=&\! -(p_d+k_d)^\mu\; \delta^{\alpha\beta} +
       p_d^\alpha \;\delta^{\mu\beta} + k_d^\beta \;\delta^{\mu\alpha} 
       + \nonumber \\ 
       & & \kappa\; (Q^\beta \; \delta^{\mu\alpha} -Q^\alpha\; \delta^{\mu\beta}). 
      \label{spin1vert}
\end{eqnarray}
The Dirac indices $\alpha,\beta$ in (\ref{spin1vert}) refer to the
vector indices of the final and initial state wave function, respectively.
The axialvector diquark can have an anomalous magnetic moment $\kappa$.
We obtain its value from a calculation for vanishing momentum transfer
($Q^2=0$) in which the quark substructure of the diquarks is
resolved, {\em i.e.}, in which a (soft) photon couples to the quarks within the
diquarks. The corresponding contributions are represented by the upper and
the right diagram in figure~\ref{emresolve}. The calculation of   $\kappa$ 
is provided in Appendix~\ref{dqres1}. 
The values obtained from the two parameter sets are both very
close to   $\kappa = 1$  (see table \ref{cc_1}). This might seem
understandable from nonrelativistic intuition: the magnetic moments of two
quarks with charges $q_1$ and $q_2$ add up to $(q_1+q_2)/m_q$, the magnetic
moment of the axialvector diquark is $(1+\kappa)(q_1+q_2)/m_{ax}$ and if the
axialvector diquark is weakly bound, $m_{ax}\backsimeq 2 m_q$, then $\kappa
\backsimeq 1$. In the following we use $\kappa=1$.

The vertices in eqs. (\ref{qandsdqv}) and (\ref{spin1vert}) satisfy 
their respective Ward-Takahashi identities, {\it i.e.} those for free
quark and diquark propagators ({\it c.f.}, eqs. (\ref{Ds},\ref{Da}) and
(\ref{S})), 
and thus describe the minimal coupling of the photon to quark and diquark.

We furthermore take into account impulse approximate contributions 
describing the photon-induced transitions \linebreak 
between  scalar and axialvector
diquarks as represented by the lower diagram in figure \ref{impulse}. 
These yield purely transverse currents and do thus not affect current
conservation. The tensor structure of these contributions resembles that of
the triangle anomaly. In particular, the structure of the 
vertex describing the transition from axialvector (with index $\beta$) to
scalar diquark is given by 
\begin{equation}
 \Gamma^{\mu\beta}_{sa}=-i\frac{\kappa_{sa}}{2M_n}\, \epsilon^{\mu\beta\rho\lambda} 
                 (p_d+k_d)^\rho Q^\lambda , 
 \label{sa_vert}
\end{equation}
and that for the reverse transition from an scalar to axialvector (index $\alpha$) by,
\begin{equation}
 \Gamma^{\mu\alpha}_{as}=i\frac{\kappa_{sa}}{2M_n}\, \epsilon^{\mu\alpha\rho\lambda}  
                 (p_d+k_d)^\rho Q^\lambda.
 \label{as_vert}
\end{equation}
The tensor structure of these anomalous diagrams (for $Q\to 0$) is derived by
resolving the diquarks in Appendix \ref{dqres1} in a way as
represented by the lower diagram in figure~\ref{emresolve}. The  explicit
factor $1/M_n$ was 
introduced to isolate a dimensionless constant $\kappa_{sa}$. Its value is 
obtained roughly as $\kappa_{sa}\simeq 2.1$ (with the next digit depending on
the parameter set, {\it c.f.} table \ref{cc_1}).

\begin{figure}[t]
\begin{center}
 \epsfig{file=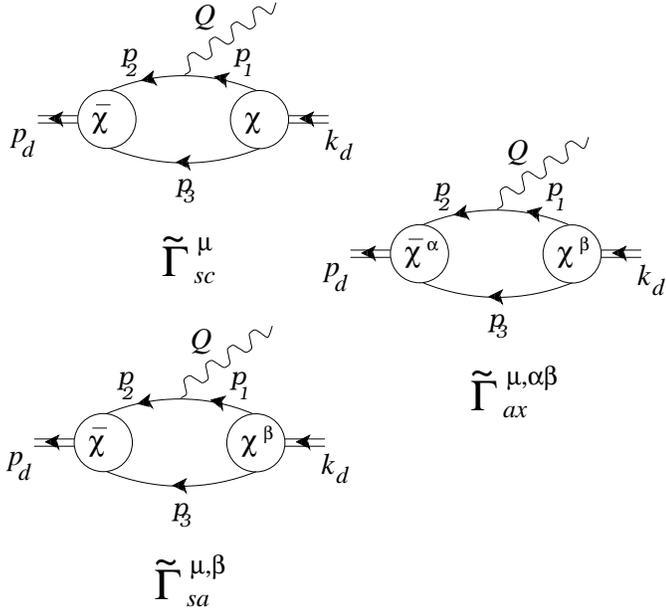, width=\columnwidth} 
\end{center}
\caption{Resolved vertices: photon-scalar diquark, photon-axialvector diquark
  and anomalous scalar-axialvector diquark transition.}
\label{emresolve}
\end{figure}

Upon performing the flavor algebra for the current matrix elements of the
impulse approximation, one obtains the following explicit forms for
proton and neutron,
\begin{eqnarray}
 \langle J^\mu \rangle^{\rm imp}_{p} &=&
   \frac{2}{3} \langle J^\mu_q \rangle^{\rm sc-sc} +
   \frac{1}{3} \langle J^\mu_{sc} \rangle^{\rm sc-sc} +
   \langle J^\mu_{ax} \rangle^{\rm ax-ax} + \nonumber \\
   & &\frac{\sqrt{3}}{3} \left( \langle J^\mu_{sa} \rangle^{\rm sc-ax} +
         \langle J^\mu_{as} \rangle^{\rm ax-sc} \right) \; , \label{jimp}\\
 \langle J^\mu \rangle^{\rm imp}_{n} &=&
   -\frac{1}{3}\left( \langle J^\mu_q \rangle^{\rm sc-sc} -
    \langle J^\mu_q \rangle^{\rm ax-ax} -
    \langle J^\mu_{sc} \rangle^{\rm sc-sc} + \right. \nonumber \\
   &&\left. \langle J^\mu_{ax} \rangle^{\rm ax-ax} \right)- 
   \frac{\sqrt{3}}{3} \left( \langle J^\mu_{sa} \rangle^{\rm sc-ax} +
         \langle J^\mu_{as} \rangle^{\rm ax-sc} \right) \; .\label{jimn}
\end{eqnarray}
The superscript `sc-sc' indicates that
the current operator is to be sandwiched between scalar
nucleon amplitudes for both the final and the initial state in
eq.~(\ref{curmelt}).
Likewise `sc-ax' denotes current operators that are sandwiched
between scalar amplitudes in the final and axialvector amplitudes in
the initial state, {\em etc.}. We note that the
axialvector amplitudes contribute to the proton current
only in combination with diquark current couplings.

Current conservation requires that the photon also be coupled to
the interaction kernel in the BS equation, {\it i.e.}, to the second term in
the inverse quark-diquark propagator $G^{-1}$ of eq.~(\ref{Kdef}). The
corresponding contributions were derived in \cite{Oet99}. They are 
represented by the diagrams in figure\ \ref{7dim}. In particular, in
\cite{Oet99} it was shown that, in addition to the photon coupling with the
exchange-quark (with vertex $\Gamma^\mu_q$), irreducible  (seagull) 
interactions of the photon with the diquark substructure have to be taken
into account. The structure and functional form of these diquark-quark-photon
vertices is constrained by Ward identities. The explicit construction of 
ref.~\cite{Oet99} yields seagull couplings of the following form
(with $M^\mu$ denoting that for the scalar diquark and $M^{\mu,\beta}$ 
that for the axialvector with Lorentz index $\beta$): 
\begin{eqnarray}
  M^{\mu[,\beta]} & = & q_q \frac{(4p_1-Q )^\mu}
                                 {4 p_1\cdot Q -Q^2} 
                 \left[ \chi^{[\beta]}( p_1-Q/2 ) -\chi^{[\beta]}( p_1) \right] 
                        + \nonumber \\
             & &   q_{ex} \frac{(4p_1+Q )^\mu}
                                 {4 p_1\cdot Q +Q^2} 
                 \left[ \chi^{[\beta]}( p_1+Q/2 ) -\chi^{[\beta]}( p_1)
                                 \right] \; . \label{SeagullsM}
\end{eqnarray}
Here, $q_q$ denotes the charge of the quark with momentum $p_q$, $q_{ex}$ the
charge  
of the exchanged quark with momentum $q'$, and $p_1$ is the relative momentum
of the two, $p_1=(p_q-q')/2$ (see figure \ref{7dim}). The conjugate vertices
$\bar M^{\mu[,\alpha]}$ are
obtained from the conjugation of the diquark amplitudes $ \chi^{[\beta]}$ in
eq.~(\ref{SeagullsM}) together with the replacement $p_1 \to p_2 = (q-k_q)/2$
({\it c.f.}, ref.~\cite{Oet99}).

\begin{figure}[t]
\begin{center}
 \epsfig{file=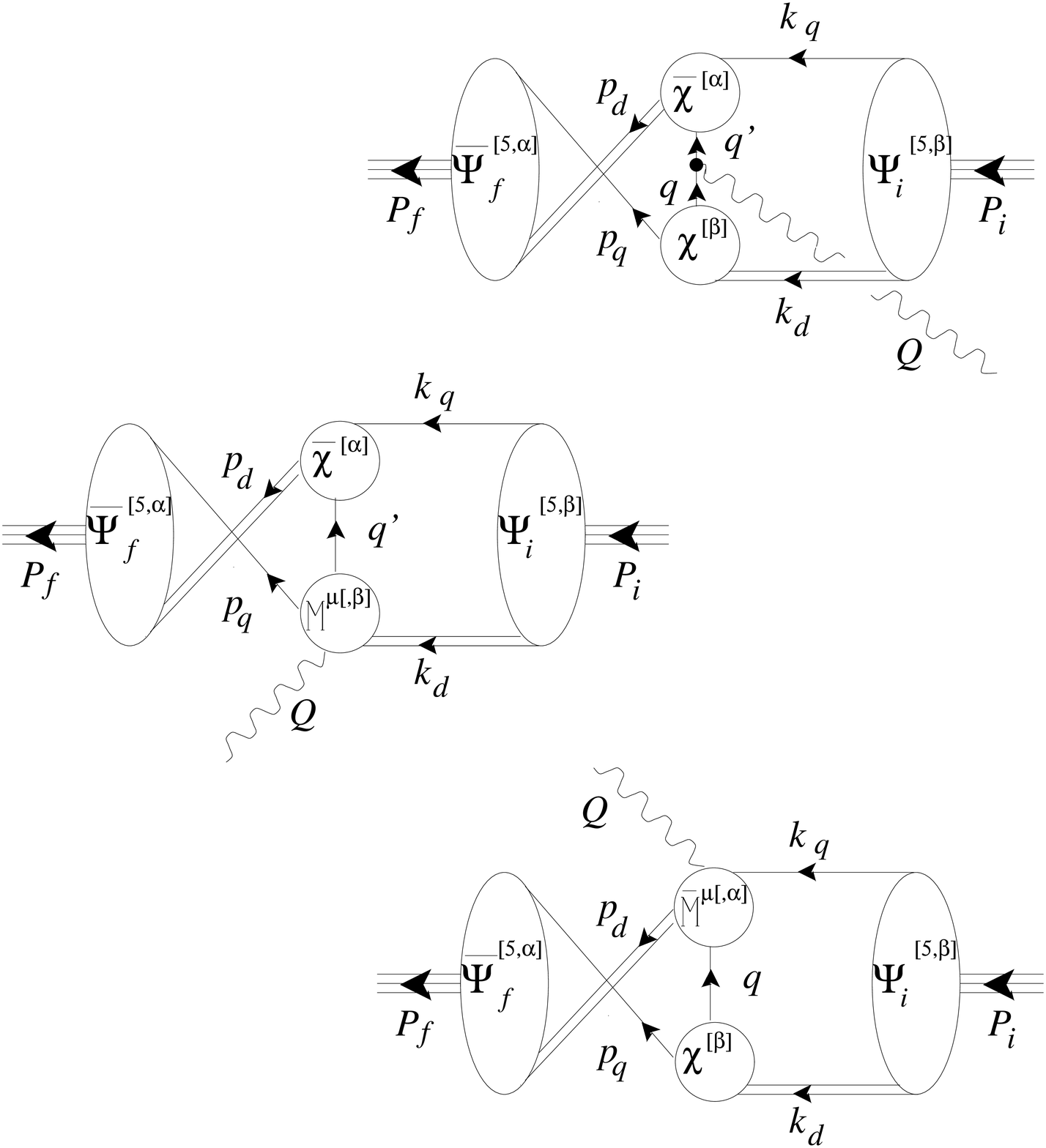,width=\columnwidth}
\end{center}
\caption{Exchange quark and seagull diagrams.}
\label{7dim}
\end{figure}

Regarding the numerical evaluation of the diagrams we remark that
in the case of the impulse approximation diagrams we
use the covariant decomposition of the vertex function $\Phi$
given in eqs.~(\ref{phi5deco},\ref{phimudeco}) 
together with the numerical solution for
the scalar functions $S_i,\,A_i,\,B_i,\,C_i$. The continuation
of these functions from the nucleon rest frame to the Breit frame is
described in detail in ref.~\cite{Oet99}. For finite momentum transfer,
care is needed in treating the singularities of the
quark and diquark propagators that appear in the single
terms of eq.~(\ref{curmelt}). In ref.~\cite{Oet99} it was shown that 
for some kinematical situations explicit residues have 
to be taken into account in the calculation of the impulse
approximation diagrams. This applies also to the calculations
presented here.

The computation of the diagrams given in figure \ref{7dim}
involves two four-dimensional integrations. As the singularity
structure of these diagrams becomes quite intricate, we
resorted to  an expansion of the wave function $\Psi$
analogously to the expansion of $\Phi$. The corresponding scalar
functions that have been  computed in the rest frame as well show a much weaker 
convergence in terms of the Chebyshev polynomials than the
scalar functions related to the expansion of the vertex function
\cite{Oet99}.
As a result, the numerical uncertainty (for these diagrams only)
exceeded the level of a few percent beyond momentum transfers
of 2.5 GeV$^2$. Due to this limitation (which could
be avoided by increasing simply the used computer time)
the form factor results presented in section \ref{results}
are restricted to the region
of momentum transfers below 2.5 GeV$^2$.

These remarks concerning the numerical method
apply also to the computation of the
strong form factor $g_{\pi NN}$ and the weak form factor
$g_A$ which will be discussed in the following subsection.

\refstepcounter{subsection} 
\label{formfac_gp}

\newpage
%\bigskip
\leftline{\it \thesubsection\  
The Strong Form Factor $g_{\pi NN}$ and the Weak Form} 
 {\it Factor $g_A$}  
\smallskip

The coupling of the pion to the nucleon, described by a pseudoscalar operator,
and the pseudovector
currents of weak processes such as the neutron $\beta$-decay are connected to
each other in the soft limit by the Goldberger-Treiman relation. 

The (spin-summed) matrix element of the pseudoscalar density $J_5^a$ can be
parameterized as 
\begin{equation}
  \langle J^a_5 \rangle = \Lambda^+(P_f)\,\tau^a \gamma_5 g_{\pi NN}(Q^2)\,
  \Lambda^+(P_i).    \label{psdens}
\end{equation}
Straightforward Dirac algebra allows to extract form factor as the
following trace,
\begin{equation}
 g_{\pi NN}(Q^2) = - \frac{2M_n^2}{Q^2} \text{Tr} \langle J^a_5 \rangle
          \gamma_5   \frac{\tau^a}{2}. \label{gpNNtr}  
\end{equation}
To compute the form factors we first specify
a suitable quark-pion vertex, evaluate an impulse
approximate contribution corresponding to the upper diagram of
figure~\ref{impulse}, and an exchange contribution analogous to the upper diagram
of figure~\ref{7dim}.
The structures and strengths (for $Q\to 0$) for the couplings of the diquarks
to the pion and the axialvector current (the remaining two impulse
approximation diagrams  in figure\ \ref{impulse}) are obtained from resolving
the diquarks in a way similar to their electromagnetic couplings in 
Appendix \ref{dqres2}.  

The structure of the inverse quark propagator, given by $S^{-1}(p)=-i\pslash
A(p^2)-B(p^2)$, suggests that we use for the 
pion-quark vertex 
\begin{equation}
 \Gamma_5%^{on-shell} 
         = -\gamma_5 \frac{B}{f_\pi} \tau^a ,
 \label{vertpion}
\end{equation}
and discard the three additionally possible Dirac structures ($f_\pi$ is
the pion decay constant). The reason is that in the chiral limit
eq.~(\ref{vertpion}) is the exact pion BS amplitude for equal quark and
antiquark momenta, since the Dyson-Schwinger equation for the scalar function
$B$ agrees with the BS equation for a pion of zero momentum in this limit. 
Of course, the subdominant amplitudes should in principle be included
for physical pions (with momentum $- P^2 = m_\pi^2$), when 
solving  the Dyson-Schwinger equation for $A$ and $B$ 
and the BS equation for the pion in mutually consistent truncations
\cite{Ben96,Mar97a,Mar97b}. Herein we employ $A(p^2)=1$ and $B(p^2)=m_q$.

The matrix elements of the pseudovector current are para\-me\-terized
by the form factor $g_A(Q^2)$ and the induced pseudoscalar form factor
$g_P(Q^2)$, 
\begin{equation}
 \langle J^{a,\mu}_5 \rangle = \Lambda^+(P_f)\,\frac{\tau^a}{2}
  \left[ i\gamma^\mu \gamma_5 g_A(Q^2) +Q^\mu \gamma_5 g_P(Q^2) \right]\,\Lambda^+(P_i).
  \label{nucax}
\end{equation}
For $Q^2 \to 0$ the Goldberger-Treiman relation, 
\begin{equation}
 g_A(0) =  f_\pi  g_{\pi NN}(0)/{M_n} \; ,
\end{equation}
then follows from current conservation and the observation 
that only the induced pseudoscalar form factor $g_P(Q^2)$ has a pole on the
pion mass-shell. 

By definition, $g_A$ describes the regular part of the pseudovector current
and $g_P$ the induced pseudoscalar form factor.
They can be extracted from eq.~(\ref{nucax}) as follows:
\begin{eqnarray}
 g_A (Q^2) & = &- \frac{i}{4\left( 1+\frac{Q^2}{4 M_n^2}\right) }
             \text{Tr} \langle J^{a,\mu}_5 \rangle \left( \gamma_5 \gamma^\mu -
             i \gamma_5 \frac{2M_n}{Q^2} Q^\mu \right) \tau^a \; ,
 \label{gatrace} \nonumber \\
           & & \\
 g_P (Q^2) & = & \frac{2M_n}{Q^2} \left( g_A (Q^2) - \frac{M_n}{Q^2}
              \text{Tr} \langle J^{a,\mu}_5 \rangle Q^\mu \gamma_5 \;\tau^a
               \right) \; .
 \label{gptrace}
\end{eqnarray} 
We again use chiral
symmetry constraints to construct the pseudovector-quark vertex.   
In the chiral limit, the Ward-Ta\-ka\-ha\-shi identity for this vertex reads, 
\begin{equation}
 Q^\mu \Gamma^{\mu,a}_5 = \frac{\tau^a}{2}
      \left( S^{-1}(k) \gamma_5 + \gamma_5 S^{-1}(p) \right)\; ,  \quad (Q=k-p).
\label{pvWTI}
\end{equation}
To satisfy this constraint we use the form of the vertex proposed in
ref.~\cite{Del79},  
\begin{equation}
 \Gamma^{\mu,a}_5 = 
       -i \gamma^\mu \gamma_5 \frac{\tau^a}{2} +  \frac{Q^\mu}{Q^2} 
       f_\pi \Gamma_5^a .
 \label{vertpv}
\end{equation}
The second term which contains the massless pion pole
does not contribute to $g_A$ as can be seen from eq.\ (\ref{gatrace}).
From these quark contributions to the pion coupling 
and the pseudovector current alone, eqs. (\ref{vertpv}) and (\ref{gpNNtr}) 
would thus yield, 
\begin{equation}
 \lim_{Q^2 \rightarrow 0} \frac{Q^2}{2M_n} g_P(Q^2) = 
   \frac{f_\pi}{M_n} g_{\pi NN} (0).  
   \label{gpgpNN}
\end{equation}
Here, the Goldberger-Treiman relation followed if the pseudovector
current was conserved or, off the chiral limit, from PCAC. 

Current conservation is a non-trivial requirement
in the relativistic bound state description of nucleons, however. 
First, we ignored the pion and pseudovector couplings to the 
diquarks in the simple argument above. For scalar diquarks alone which
themselves do not couple to either of the two,\footnote{This can be inferred
from parity and covariance.}
pseudovector current conservation could in principle 
be maintained by including the couplings to the interaction kernel of the
nucleon BS equation in much the same way as was done for the electromagnetic
current. 

Unfortunately, when axialvector diquarks are included, 
even this will not suffice to maintain current conservation.
As observed recently in ref.~\cite{Ish00}, a doublet of axialvector {\em and}
vector diquarks has to be introduced, in order to comply with chiral
Ward identities in general. The reason essentially is that vector and
axialvector diquarks mix under a chiral
transformation whereas this is not the case
for scalar and pseudoscalar diquarks. 
Since vector diquarks on the one hand 
introduce six additional components to the nucleon wave function, 
but are on the other hand not expected to influence the binding strongly,
here we prefer to neglect vector diquark correlations and to
investigate the axial form factor without them. 

The pion and the pseudovector current can couple to the diquarks
by an intermediate quark loop. As for the anomalous contributions to the 
electromagnetic current, we derive the Lorentz structure of the 
diquark vertices and calculate their effective strengths from this
quark substructure of the diquarks in Appendix\ \ref{dqres2}. 

As mentioned above, no such couplings arise for the scalar diquark.
The axialvector diquark and the pion couple by an anomalous vertex.
Its Lorentz structure is similar to that for the photon-induced
scalar-to-axialvector transition in eq.\ (\ref{sa_vert}): 
\begin{eqnarray}
 \Gamma^{\rho\lambda,abc}_{5,ax} &=& \frac{\kappa^5_{ax}}{2M_n} 
                      \frac{m_q}{f_\pi}  
                     \epsilon^{\rho\lambda\mu\nu}
                  (p_d+k_d)^\mu Q^\nu\, (1-2\delta^{a2})
                    i\epsilon^{abc}
                 \label{vert5ax}.
\end{eqnarray}
Here,  $a$ is the flavor index of the pion or, below, of the pseudovector
current, while $b$ and $c$ are those of the outgoing and the incoming
axialvector diquarks (with Lorentz indices $\rho$ and $\lambda$) according to
eq.\ (\ref{dqvertex_a}), respectively.  
The factor $m_q/f_\pi$ comes from quark-pion vertex (\ref{vertpion})
in the quark-loop (see Appendix\ \ref{dqres2}), and the nucleon mass
was introduced to isolate a dimensionless constant $\kappa^5_{ax}$.

The pseudovector current and the axialvector diquark are also coupled by 
anomalous terms. As before, we denote with $\rho$ and $\lambda$ the Lorentz
indices of outgoing and incoming diquark, respectively,
and with $\mu$ the pseudovector index. Out of three possible Lorentz
structures for the regular 
part of the vertex, $p_d^\mu \;\epsilon^{\rho\lambda\alpha\beta} p_d^\alpha
Q^\beta$, $\epsilon^{\mu\rho\lambda\alpha}Q^\alpha$ and
$\epsilon^{\mu\rho\lambda\alpha} (p_d+k_d)^\alpha$, in the limit $Q
\rightarrow 0$ only the last term contributes to $g_A$. 
We furthermore verified numerically that the first two terms yield negligible
contributions to the form factor also for finite $Q$. Again, the pion pole
contributes proportional to $Q^\mu$, and our ansatz for the vertex thus reads
\begin{eqnarray}
 \Gamma^{\mu\rho\lambda,abc}_{5,ax} &=& \frac{\kappa^5_{\mu,ax}}{2} 
                        \epsilon^{\mu\rho\lambda\nu} (p_d+k_d)^\nu \;
                    \frac{1}{2} (1-2\delta^{a2})
                    i\epsilon^{abc} + \nonumber \\
                   & &\frac{Q^\mu}{Q^2} 
                    f_\pi \Gamma^{\rho\lambda,abc}_{5,ax} \label{vert5muax}.
\end{eqnarray}
For both strengths we roughly obtain  
$\kappa^5_{ax} \simeq \kappa^5_{\mu,ax} \simeq 4.5$ slightly dependent on the
parameter set, see table~\ref{cc_2}  (in Appendix \ref{dqres2}).  

Scalar-to-axialvector transitions are also possible by the pion and the
pseudovector current. An effective vertex for the pion-mediated transition
has one free Lorentz index to be contracted with the axialvector
diquark. Therefore, two types of structures exist, one with the pion momentum
$Q$, and the other with any combination of the diquark momenta $p_d$ and
$k_d$. If we considered this transition as being described by an interaction 
Lagrangian of scalar, axialvector and pseudoscalar fields, terms of the
latter structure were proportional to the divergence of the axialvector field
which is a constraint that can be set to zero.
We therefore adopt the following form for the transition vertex,
\begin{equation}
 \Gamma^{\rho,ab}_{5,sa} = -i \kappa^5_{sa} 
                     \frac{m_q}{f_\pi}  
                    Q^\rho \; (2\delta^{a2}-1)\delta^{ab}.
 \label{vert5sa}
\end{equation} 
The flavor and Dirac indices of the axialvector diquark are $b$ and $\rho$.
This vertex corresponds to a derivative coupling of the
pion to scalar and axialvector diquark.

The pseudovector-induced transition vertex has two Lorentz indices,
denoted by $\mu$ for the pseudovector current and $\rho$ for the axialvector
diquark. 
From the momentum transfer $Q^\mu$ and one of the diquark momenta,
altogether five independent tensors can be constructed:
$\delta^{\mu\rho}, \;Q^\mu Q^\rho, \;Q^\mu p_d^\rho, \;p_d^\mu Q^\rho$
and $p_d^\mu p_d^\rho$ (the totally antisymmetric tensor 
has the wrong parity).
We assume as before, that all terms proportional to $Q^\mu$ are contained
in the pion part (and do not contribute to $g_A$). From the
diquark loop calculation in Appendix\ \ref{dqres2} we find that
the terms proportional to $p_d^\mu Q^\rho$ and $p_d^\mu p_d^\rho$
can again be neglected with an error on the level
of one per cent. Therefore, we use a vertex of the form,
\begin{equation}
 \Gamma^{\mu\rho,ab}_{5,sa} = i M_n \kappa^5_{\mu,sa}\;  
                            \delta^{\mu\rho}
 \;\frac{1}{2}(2\delta^{a2}-1)\delta^{ab}
                    + \frac{Q^\mu}{Q^2} f_\pi \Gamma^{\rho,ab}_{5,sa}.
 \label{vert5musa}
\end{equation}
For the strengths of these two transition vertices 
we obtain $\kappa^5_{sa}\simeq 3.9$ and (on average)
$\kappa^5_{\mu,sa} \simeq 2.1$, see table~\ref{cc_2} in Appendix\
\ref{dqres2}.  

We conclude this section with 
the observation that eq.~(\ref{gpgpNN}) remains valid with   
{\em all} contributions from quarks {\em and} diquarks included
in the pseudoscalar density (\ref{psdens}) and the pseudovector current
(\ref{nucax}) of the nucleon. The extension of this statement to include the
diquark couplings follows from the fact that their vertices for the
pseudovector current, eqs.~(\ref{vert5muax}) and (\ref{vert5musa}), 
contain the pion pole in the form entailed by their Ward identities.
Eq.~(\ref{gpgpNN}) is then obtained straightforwardly by 
inserting  $g_{\pi NN}$ from eq.~(\ref{gpNNtr}) term by term into 
the corresponding equation (\ref{gptrace}) for $g_P$ and expanding 
to leading order in $Q^2$.

\section{Results}
\label{results}
%\subsection{Electromagnetic Form Factors}

\begin{figure*}[t]
 \begin{center}
  \epsfig{file=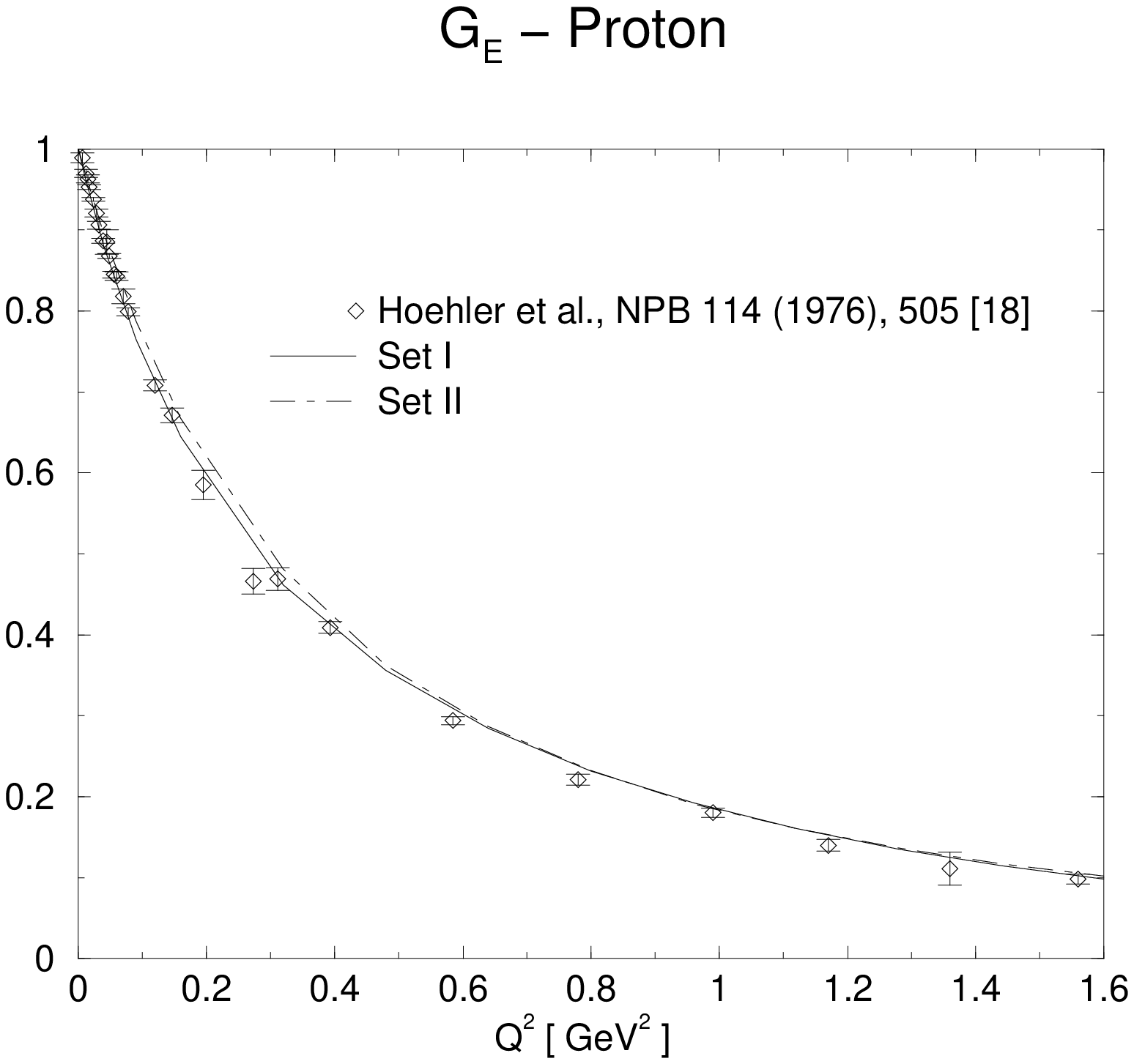,width=7cm} \hspace{1cm}
  \epsfig{file=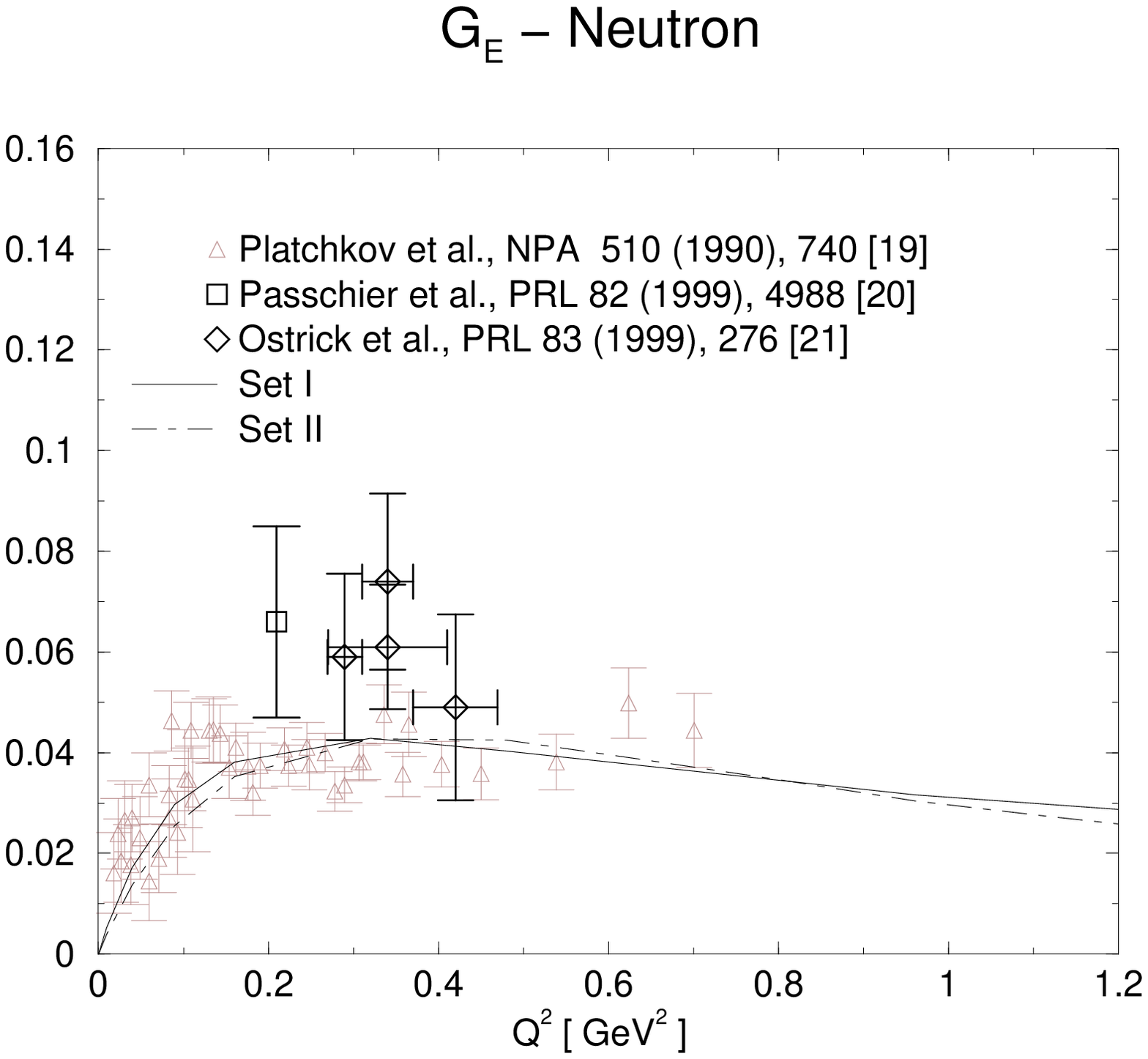,width=7cm}
 \end{center}
 \caption{Electric form factors of both nucleons for the parameter
  sets of table~\ref{pars}. Experimental data
  for the proton is taken from \cite{Hoe76}. The older neutron data
  analysis \cite{Pla90} contains more systematic uncertainties (due to
  specific nucleon-nucleon potentials) than the more recent data from
  \cite{Pas99,Ost99}.} 
 \label{gefig}  
\end{figure*}

\refstepcounter{subsection}
\leftline{\it 4.1 Electromagnetic Form Factors}
\smallskip

The results for the electric form factors are shown in figure\ \ref{gefig}.
The phenomenological dipole behavior of the proton $G_E$ is well 
reproduced by both parameter sets.
The electric radius (see table\ \ref{radtab}) 
is predominantly sensitive to the width of the BS wave function.
 This is different for the neutron. Here, stronger axialvector
correlations tend to suppress the electric form factor as compared to the 
calculation of ref.\ \cite{Oet99} where only scalar diquarks  were
maintained. 
A smaller binding energy compensates this effect, so that the results for    
$G_E$ are basically the same for both parameter sets, see figure \ref{gefig}.

\begin{table}[b]
 \begin{center}
  \begin{tabular}{ll|rr|r}
 &  & Set I & Set II  & expt. \\
 & & & & \\ \hline
 $(r_p)_{\rm el}$ & [fm]  & 0.88 & 0.81 & $0.836\pm 0.013$ \\ 
 $(r^2_n)_{\rm el}$ & [fm$^2$] & $-$0.12 & $-$0.10 & $-0.113 \pm 0.007$ \\
 $(r_p)_{\rm mag}$  & [fm] & 0.84  & 0.83 & $0.843 \pm 0.013$   \\
 $(r_n)_{\rm mag}$  & [fm] & 0.84 & 0.83  & $0.840 \pm 0.042$ \\
 &  & & & \\
  \end{tabular}

  \caption{Nucleon electric and magnetic radii for the two parameter sets
  compared to the experimental values from \cite{Kop95} 
  (for $(r^2_n)_{\rm el}$) and \cite{Hoe76} (for the remaining radii).} 
  \label{radtab}
 \end{center} 
\vskip -.5cm
\end{table}

For the magnetic moments summarized in table\ \ref{magtab} two parameters
are important. First, in contrast to non-relativistic constituent models,
the dependence of the proton magnetic moment on the ratio $M_n/m_q$  
is stronger than linear. As a result, the quark impulse contribution to
$\mu_p$ with the scalar diquark being spectator, which is the dominant one, 
yields about the same for both sets, even though the 
corresponding nucleon amplitudes of Set I contribute about 25\% less 
to the norm than those of Set II. Secondly, the scalar-axialvector
transitions contribute equally strong (Set I) or stronger (Set II)
than the spin flip of the axialvector diquark itself.
While for Set II (with weaker axialvector diquark correlations) 
the magnetic moments are about 30\% too small,  
the stronger diquark correlations of Set I yield an isovector contribution
which is only 15\% below and an isoscalar magnetic
moment slightly above the phenomenological value.

\begin{table}[b]
\begin{center}
\begin{tabular}{ll|rr|r}
   &  & Set I & Set II & expt. \\
   &  &  &  &  \\ \hline
   & Sc-q & 1.35  & 1.33 &  \\ 
   & Ax-dq & 0.44 & 0.08 & \\
  $\mu_p$ & Sc-Ax & 0.43 & 0.24 & \\
   & Ex-q & 0.25 & 0.22 &\\
   & SUM  & 2.48 & 1.92 & 2.79\\ \hline
 $\mu_n$  & SUM  & -1.53 & -1.35 & -1.91 \\ \hline
 $\mu_p + \mu_n$   & isoscalar & 0.95  & 0.57 & 0.88 \\
 $\mu_p - \mu_n$  &  isovector & 4.01  & 3.27 & 4.70 
\end{tabular}

\end{center}
\caption{Magnetic moments of proton and neutron. 
For the proton we list the following contributions separately: 
from the impulse quark-coupling with scalar nucleon amplitudes 'Sc-q', from 
the axialvector diquark 'Ax-dq', from the scalar-axialvector transition
'Sc-Ax', and from  the exchange quark 'Ex-q'.
Seagull and scalar diquark contributions are small.}
\label{magtab}
\end{table}

Stronger axialvector diquark correlations are favorable for larger values of
the magnetic moments as expected. If the isoscalar magnetic moment is taken
as an indication that those of Set I are somewhat too strong, however, 
a certain mismatch with the isovector contribution remains, also with 
axialvector diquarks included.

Recent data from ref.~\cite{Jon99} for the ratio $\mu_p\, G_E/G_M$ is 
compared to our results in figure\ \ref{gemfig}. The
ratio obtained from Set II with weak axialvector correlations
lies above the experimental data, and that for Set I below.
The experimental observation that this ratio decreases significantly with
increasing $Q^2$ (about 40\% from $Q^2 = 0 $ to $3.5$ GeV$^2$), can be
well reproduced with axialvector diquark correlations of a certain strength
included. The reason for this is the following:
The impulse approximate photon-diquark couplings yield contributions that 
tend to fall off slower with increasing $Q^2$ than those of the quark. 
This is the case for both, the electric and the magnetic form factor. 
If no axialvector diquark correlations inside the nucleon are maintained,
however, the only diquark contribution to the electromagnetic current arises
from $\langle J^\mu_{sc} \rangle^{\rm sc-sc}$, 
see eqs.~(\ref{jimp},\ref{jimn}). Although this term does provide for
a substantial contribution to $G_E$, its respective contribution
to $G_M$ is of the order of 10$^{-3}$. This reflects the fact that
an on-shell scalar diquark would have no magnetic moment at all, and
the small contribution to $G_M$ may be interpreted as an off-shell effect. 
Consequently, too large a ratio $\mu_p\, G_E/G_M$ results, if only scalar
diquarks are maintained \cite{Oet99}. For Set II (with weak axialvector
correlations), this effect is still visible, 
although the scalar-to-axialvector transitions
already bend the ratio towards lower values at larger $Q^2$. These transitions
almost exclusively contribute to $G_M$, and it thus follows that the stronger
axialvector correlations of Set I enhance this effect. 
Just as for the isoscalar magnetic moment, the axialvector diquark
correlations of Set I tend to be somewhat too strong here again, however.     
To summarize, 
the ratio $\mu_p \, G_E/G_M$ imposes an upper limit on the relative importance 
of the axialvector correlations of estimated 30\% (to the BS norm of the
nucleons).  
This finding will be confirmed once more in our analysis of the pion-nucleon
and the axial coupling constant below.

%\newpage

\begin{table}[b]

\begin{tabular}{l|rr|rr}
   &  \multicolumn{2}{c}{Set I}  &  \multicolumn{2}{c}{Set II}   \\   
   &  $g_{\pi NN}(0)$ & $g_A(0)$ & $g_{\pi NN}(0)$ & $g_A(0)$  \\  
   &  &  &  &  \\  \hline  
 Sc-q  & 7.96 & 0.76 & 9.25 & 0.86  \\   
 Ax-q  & 0.50 & 0.04 & 0.10 & 0.01 \\   
 Ax-dq  & 1.44 & 0.18 & 0.34 & 0.04 \\   
 Sc-Ax  & 5.66 & 0.39 & 3.79  & 0.22 \\   
 Ex-q  & 1.69 & 0.12 & 2.70 & 0.22 \\   
 SUM  & 17.25 & 1.49 & 16.18 & 1.35  \\ \hline  
 expt.  & \multicolumn{2}{r|}{$g_{\pi NN}:$ $13.14\pm 0.07$ \cite{Arn94}} & 
          \multicolumn{2}{r}{$g_{a}:$ $1.267 \pm 0.0035$ \cite{Rev98}} \\
   & \multicolumn{2}{r|}{$13.38\pm 0.12$ \cite{Rah98}} &  &       
\end{tabular}
\caption{Various contributions to $g_{\pi NN}(0)$ and $g_A(0)$, labelled as in
   table~\ref{magtab} (with 'Ax-q' for the impulse quark-coupling with
axialvector amplitudes).}
\label{ga&gpitab}   
\end{table}

%\subsection{$g_{\pi NN}$ and $g_A$}

\refstepcounter{subsection}
\bigskip
\leftline{\it 4.2 $g_{\pi NN}$ and $g_A$}
\smallskip

Examining $g_{\pi NN}(0)$ which is assumed to be close to 
the physical pion-nucleon coupling at $Q^2=-m_\pi^2$ (within 10\% by PCAC),
and the axial coupling constant $g_A(0)$, we find large contributions to
both of these arising from the scalar-axialvector transitions, see table\
\ref{ga&gpitab}. As mentioned  
in the previous section, the various diquark contributions violate
the Goldberger-Treiman relation. 
Some compensations occur between the small contributions from
the axialvector diquark impulse-coupling and the comparatively large 
ones from scalar-axialvector transitions which provide for the dominant effect
to yield %\linebreak \mbox{ $g_A(0) > 1$.} 
$g_A(0) > 1$.

\begin{table}[b]
 \begin{center}
 \begin{tabular}{ll|ll|l}
  & & Set I & Set II & expt. \\
  & & & & \\ \hline 
 $r_{\Sc\pi NN}$ & [fm] & 0.83 & 0.81 &  \\
 $r_A$ & [fm] & 0.82 & 0.81 & 0.70$\pm$0.09 
 \end{tabular}
 \caption{Strong radius $r_{\Sc\pi NN}$ and weak radius 
         $r_A$, the experimental value of the latter 
         is taken from \cite{Kit90}.}
 \end{center}
\label{radtab2}
\end{table}

Summing all these contributions, the Goldberger-Treiman discrepancy,
\begin{equation}
  \Delta_{GT} \equiv \frac{g_{\pi NN}(0)}{g_A(0)}\,\frac{f_\pi}{M_n}-1
\end{equation}
amounts to 0.14 for Set I and 0.18 for Set II. The larger discrepancy for Set
II (with weaker axialvector correlations) is due to the
larger violation of the Goldberger-Treiman relation from the
exchange quark contribution in this case. This contribution 
is dominated by the scalar amplitudes,
and its Goldberger-Treiman violation should therefore
be compensated by appropriate {\em chiral seagulls}. 
These discrepancies, and the overestimate of the pion-nucleon coupling, 
indicate that axialvector diquarks inside nucleons are likely to 
represent quite subdominant correlations.

%\newpage
\begin{figure}[t]
 \begin{center}
 \epsfig{file=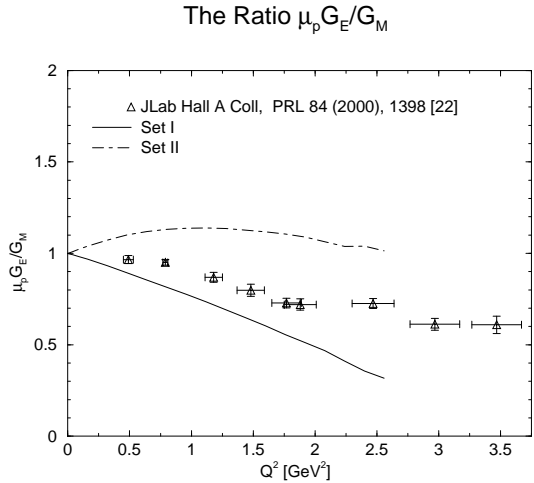,width=7cm}
 \caption{The ratio $(\mu_p\;G_E)/G_M$ compared to the data
          from ref. \cite{Jon99}.}
 \label{gemfig}
 \end{center}
\end{figure}

The strong and weak radii are presented in table\ \ref{radtab2} and the
corresponding form factors in figure\ \ref{gpifig}. The 
axial form factor is experimentally known much less precisely than 
the electromagnetic form factors. In the right panel
of figure\ \ref{gpifig}  the experimental situation is summarized
by a band of dipole parameterizations of $g_A$
that are consistent with a wide-band neutrino experiment  
\cite{Kit90}. 
Besides the slightly too large values obtained for $Q^2 \to 0$ 
which are likely to be due to the PCAC violations of axialvector diquarks as
discussed in the previous section (and which are thus less significant for
the weaker axialvector correlations of Set II), our results yield quite
compelling agreement with the experimental bounds.

\begin{figure*}[t]
 \begin{center}
   \epsfig{file=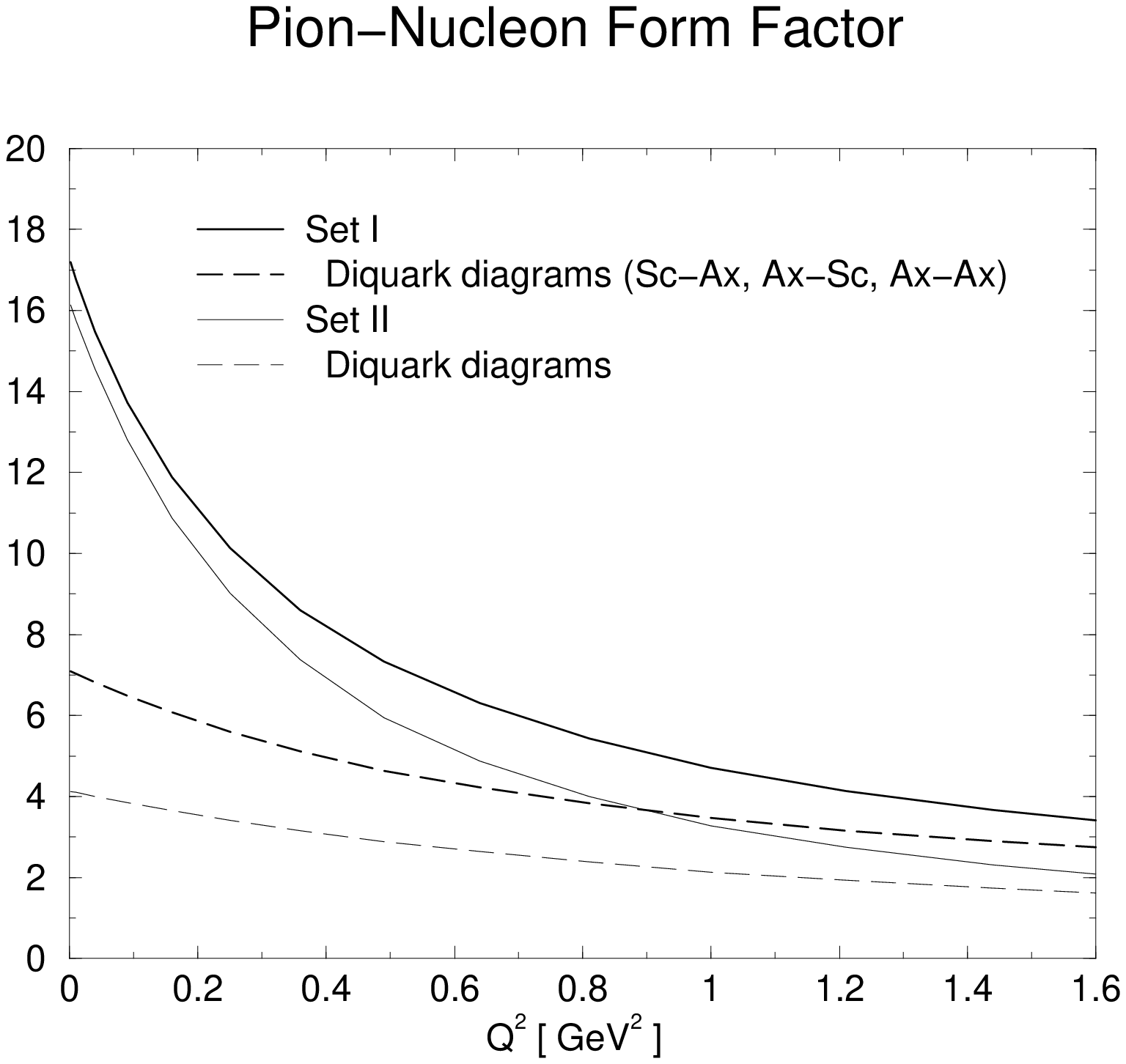,width=7cm} \hspace{1cm}
   \epsfig{file=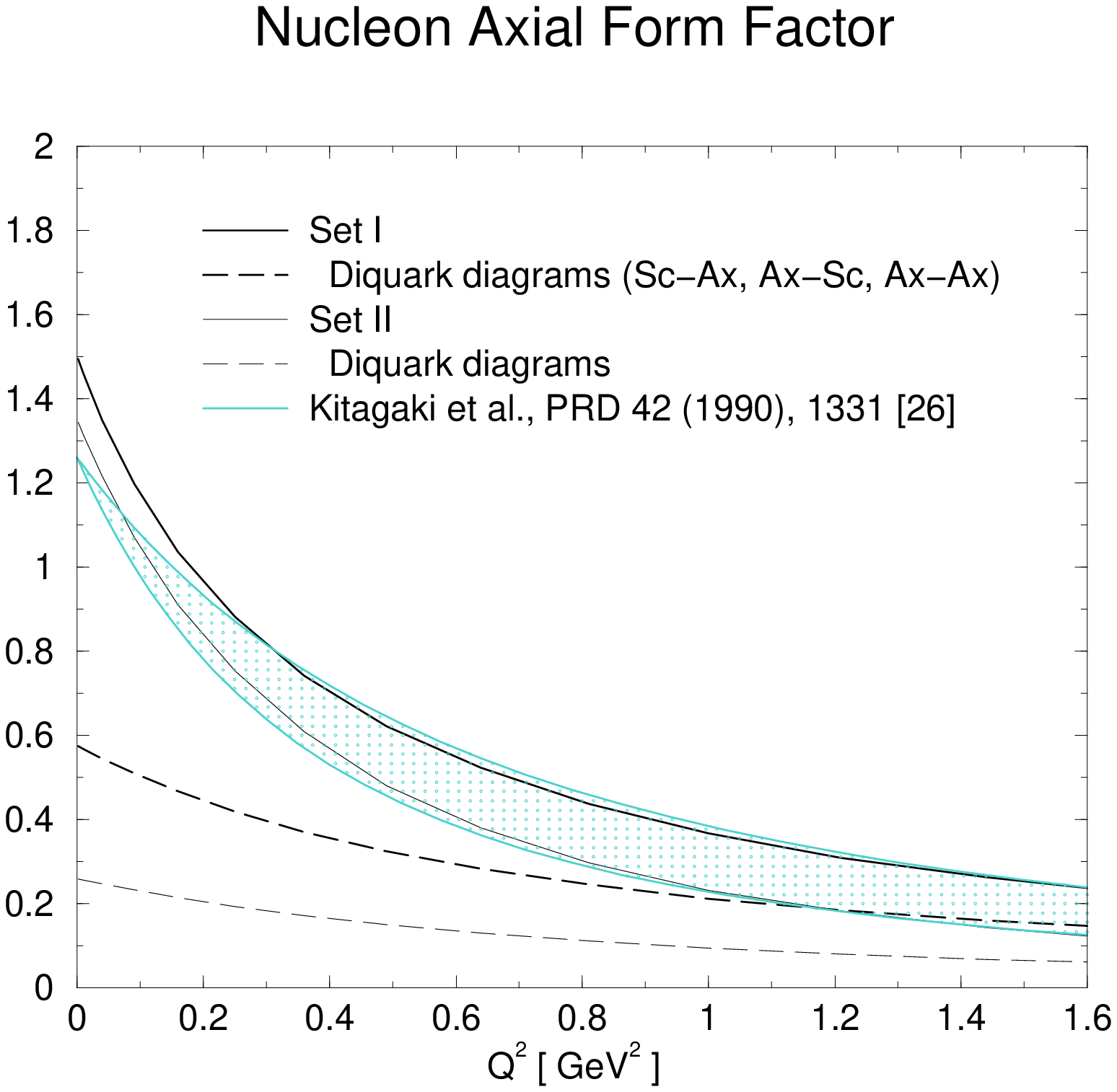,width=7cm}
 \end{center}
\caption{The pion-nucleon form factor $g_{\pi NN}(Q^2)$ and
 the axial form factor $g_A(Q^2)$. 'Diquark diagrams'
 labels the sum of the impulse approximate axialvector contributions 
 and scalar-axialvector transitions. The shaded
 region in the right panel represents the uncertainty in $g_A$ as determined
 from quasi-elastic neutrino scattering when a dipole form is fitted to both,
 the vector and the axial form factor \cite{Kit90}.}
 \label{gpifig}
\end{figure*}

\section{Summary and Conclusions}

The description of baryons as fully relativistic bound states of quark and
glue reduces to an effective Bethe-Salpeter (BS) equation with quark-exchange
interaction when irreducible 3-quark interactions are neglected and
separable 2-quark (diquark) correlations are assumed.
By including axialvector diquark correlations with non-trivial quark
substructure, we \linebreak 
solved the BS equations  of this covariant quark-diquark
model for nucleons and the $\Delta$-resonance.
While the $\Delta$ cannot be described without axialvector diquarks, the
nucleon-$\Delta$ mass splitting imposes an upper bound on their relative
importance inside nucleons, as compared to the scalar diquark
correlations. At present, this bound seems 
somewhat too strong for a simultaneous description of octet
and decuplet baryons in a fully satisfactory manner. 

We furthermore extended previous studies of nucleon properties within the 
covariant quark-diquark model. In this way we assess the influence of the
axialvector diquark correlations with non-trivial quark substructure. 
Electromagnetic form factors, the weak form factor $g_A$ and
the strong form factor $g_{\pi NN}$ have been computed.
Structures and strengths of the otherwise unknown axialvector diquark
couplings, and of scalar-axialvector transitions, have thereby been obtained
by resolving the quark-loop substructure of the diquarks at vanishing  
momentum transfer ($Q^2 \to 0$).   

An excellent description is obtained for the electric form factors of both
nucleons. The ratio of the proton electric and magnetic moment, $\mu_p \,
G_E/G_M $ as recently measured at TJNAF~\cite{Jon99}, is well described with  
axialvector diquark correlations of moderate strength. Our results clearly 
indicate that axialvector diquarks are necessary to reproduce the 
qualitative behavior of the experimental data for this ratio. At the same
time, an upper bound on the relative importance of axialvector diquarks and
scalar-axialvector transitions
(together of estimated 30\% to the BS norm of the nucleons) can be inferred. 

For axialvector correlations of such a strength, the phenomenological 
value for the isoscalar magnetic moment of the nucleons is well reproduced. 
The isovector contribution results around 15\% too small. While the
axialvector diquarks lead to a considerable improvement, both magnetic
moments tend to be around 50\% too small with scalar diquarks
alone~\cite{Oet99}, this remaining 15\% mismatch in the isovector magnetic
moment seems to be due to other effects. One possibility might be provided by
vector diquarks. While their contributions
to the binding energy of the nucleons are expected to be negligible, the
photon couplings with vector diquarks could be strong enough to compensate
this and thus lead to sizeable effects, in particular, in the magnetic
moments.   
  
For the pion-nucleon and the axial coupling constant, we found 
moderate violations of PCAC and the Goldberger-Trei\-man relation.
For scalar diquarks alone, this is attributed to some violations of 
the (partial) conservation of the simplified axial current neglecting
exchange quark couplings and chiral seagulls. Maintaining axialvector
diquarks, additional PCAC violations can arise from missing vector diquarks
which mix with the axialvectors under chiral transformations as
pointed out in ref.~\cite{Ish00}. This explains why the weaker axialvector 
correlations lead to better values for $g_A(0)$ and $g_{\pi  NN}(0)$. 
Nevertheless, these violations are reasonably small and occur only at small
momentum transfers $Q^2$. The axial form factor $g_A(Q^2)$ is otherwise in
good agreement with the experimental bounds from quasi-elastic neutrino
scattering in~\cite{Kit90}. 

It should be noted that qualitatively the scalar-axialvector transitions are
of particular importance to obtain values for $g_A > 1$. These transitions
thus solve a problem that previous applications of the covariant quark-diquark 
model shared with many chiral nucleon models. Their quantitative effect 
is somewhat too large, as discussed above. 

The conclusions from our present study can be summarized as follows:
While selected observables, sensitive to axialvector diquark correlations, 
can be improved considerably by their inclusion, other observables (and 
the nucleon-$\Delta$ mass splitting) provide upper bounds on their relative
importance as compared to scalar diquarks. These bounds confirm that
scalar diquarks provide for the dynamically dominant 2-quark
correlations inside nucleons. Deviations of the order of 15\% remain
in the isovector part of the magnetic moment (too small), in $g_A(0)$ (too
large) and in the Goldberger-Treiman relation. While these cannot be fully
accounted for by including the axialvector diquark correlations,
overall, however, the quark-diquark model was demonstrated to 
describe nucleon properties quite successfully.

\section*{Acknowledgements}

The authors gratefully acknowledge valuable discussions with 
S.~Ahlig, N.~Ishii, C.~Fischer  and  H.~Reinhardt.
The work of M.O. was supported by COSY under contract 41376610
and by the Deutsche For\-schungsge\-mein\-schaft 
under contract DFG We 1254/4-2. He is indebted to H.~Weigel for his 
continuing support.

\newpage
\begin{appendix}

\section{Resolving Diquarks}

%\subsection{Electromagnetic Vertices}

\refstepcounter{subsection}
\label{dqres1}
\leftline{\it \thesubsection\  Electromagnetic Vertices}
\smallskip

Here we adopt an impulse approximation to couple the photon directly 
to the quarks inside the diquarks obtaining the scalar, axialvector and the 
photon-induced scalar-axialvector diquark transition couplings as represented
by the 3 diagrams in figure~\ref{emresolve}. For on-shell diquarks these
yield diquark form factors and, under some mild assumptions on the
quark-quark interaction kernel~\cite{Oet99}, current conservation 
followed for amplitudes which solve a diquark BS equation.       
Due to the quark-exchange antisymmetry of the diquark amplitudes 
it suffices to calculate one diagram for each of the three contributions,
{\it i.e.}, those of figure~\ref{emresolve} in which the photon couples to the
``upper'' quark line. The color trace yields one
as in the normalization integrals, eqs. (\ref{normsc}) and (\ref{normax}).
The traces over the diquark flavor matrices with the charge matrix will be
included implicitly in those over the Dirac structures in the resolved vertices
which read (with the minus sign for fermion loops),
\begin{eqnarray}
 \tilde \Gamma^\mu_{sc} &=  -\text{Tr} \fourint{q} & 
    \bar\chi \left( \frac{\Sc p_2-p_3}{\Sc 2} \right) 
   S(p_2)(-i\gamma^\mu)S(p_1) \; \times  \nonumber \\
           & & \chi \left( \frac{\Sc p_1-p_3}{\Sc 2} \right)
    S^T(p_3)\; ,   \label{ressdqv} \\ 
 \tilde \Gamma^{\mu,\alpha\beta}_{ax} &= 
             -\text{Tr} \fourint{q} & 
      \bar\chi^\alpha \left( \frac{\Sc p_2-p_3}{\Sc 2} \right)
      S(p_2)(-i\gamma^\mu)S(p_1) \; \times \nonumber \\
           & & \chi^\beta \left( \frac{\Sc p_1-p_3}{\Sc 2}
        \right) S^T(p_3) \; ,  \label{resadqv} \\
 \tilde \Gamma^{\mu,\beta}_{sa}  &= -\text{Tr} \fourint{q} &
      \bar\chi \left( \frac{\Sc p_2-p_3}{\Sc 2} \right)
      S(p_2)(-i\gamma^\mu)S(p_1) \; \times \nonumber \\ 
          & & \chi^\beta \left( \frac{\Sc p_1-p_3}{\Sc 2} 
        \right) S^T(p_3)  \label{restv} \\
                      &=
        2im_q \epsilon^{\mu\beta\rho\lambda} & (p_d+k_d)^\rho 
                Q^\lambda \; \times \\
                 & &        \fourint{q} \frac{g_s g_a \;P(q-Q/4)P(q+Q/4)}
                        {(p_1^2+m_q^2)(p_2^2+m_q^2)(p_3^2+m_q^2)} \;
    . \nonumber  
\end{eqnarray}
The quark momenta herein are,
\begin{eqnarray}
 p_1&=&\frac{p_d+k_d}{4}-\frac{Q}{2}+q\; ,\quad
 p_2=\frac{p_d+k_d}{4}+\frac{Q}{2}+q\;,\quad \nonumber \\
 p_3&=&\frac{p_d+k_d}{4}-q \;. 
\end{eqnarray}
Even though current conservation can be maintained with these vertices
on-shell, off-shell $\tilde \Gamma^\mu_{sc}$ and $\tilde
\Gamma^{\mu,\alpha\beta}_{ax}$ do not satisfy the Ward-Takahashi identities
for the free propagators in \linebreak 
eqs.\ (\ref{Ds},\ref{Da}). They can thus not directly be
employed to couple the photon to the diquarks inside the nucleon without
violating gauge invariance. 
For $Q=0$, however, they can be used to estimate the anomalous magnetic
moment $\kappa$ of the axialvector diquark and the strength of the
scalar-axialvector transition, denoted by $\kappa_{sa}$ in (\ref{sa_vert}),
as follows:  

First we calculate the contributions of the scalar and axialvector diquark 
to the proton charge, {\it i.e.} the second diagram in figure\ \ref{impulse}
to  $G_E (0)$, with replacing the vertices $\Gamma^\mu_{sc}$ and
$\Gamma^{\mu,\alpha\beta}_{ax}$ given in eqs. (\ref{qandsdqv}) and
(\ref{spin1vert}) by the resolved ones, $\tilde \Gamma^\mu_{sc}$ and $\tilde
\Gamma^{\mu,\alpha\beta}_{ax}$ in eqs.~(\ref{ressdqv}) and (\ref{resadqv}).
Since the bare vertices on the other hand satisfy the Ward-Takahashi
identities, and since current conservation is maintained in the calculation
of the electromagnetic form factors, the correct charges of both nucleons are
guaranteed to result from the contributions to $G_E (0)$ obtained with these
bare vertices, $\Gamma^\mu_{sc}$ and $\Gamma^{\mu,\alpha\beta}_{ax}$ of
eqs.~(\ref{qandsdqv}) and~(\ref{spin1vert}). In order to reproduce the
these correct contributions, we then adjust the values for the
diquark couplings, $g_s$ and $g_a$, to be used in connection with the
resolved vertices of eqs.~(\ref{ressdqv}) and (\ref{resadqv}). This yields
couplings $g_s^{resc}$ and $g^{resc}_a$, slightly rescaled (by a factor of
the order of one, see table~\ref{cc_1}). 
Once these are fixed we can continue and calculate the contributions to the
magnetic moment of the proton that arise from the resolved axialvector and
transition couplings, $\tilde \Gamma^{\mu,\alpha\beta}_{ax}$ and
$\tilde \Gamma^{\mu,\beta}_{sa}$, respectively. These contributions determine
the values of the constants $\kappa$ and $\kappa_{sa}$ for the couplings in
eqs.~(\ref{spin1vert}) and (\ref{sa_vert}), (\ref{as_vert}). 

The results are given in table\ \ref{cc_1}.
As can be seen, the values obtained for 
$\kappa$ and $\kappa_{sa}$ by this procedure are insensitive to the parameter
sets for the nucleon amplitudes. 
In the calculations of observables we use $\kappa=1.0$ and $\kappa_{sa}=2.1$.

\begin{table}[t]
 \begin{center}
  \begin{tabular}{l|llll}
    & $g_s^{resc}/g_s$ & $g_a^{resc}/g_a$ & $\kappa$ & $\kappa_{sa}$ \\ \hline
    & & & & \\
    Set I & 0.943 & 1.421  & 1.01 & 2.09\\
    Set II& 0.907  & 3.342  & 1.04 & 2.14
  \end{tabular}\\[+4pt]
 \caption{
   Rescaled diquark normalizations 
   and constants of photon-diquark couplings.}
  \label{cc_1}
 \end{center}
 \vspace{-0.2cm}
\end{table}

%\vspace{.2cm}

%\subsection{Pseudoscalar and Pseudovector Vertices}

\refstepcounter{subsection}
\label{dqres2}
\bigskip
\leftline{\it \thesubsection\ Pseudoscalar and Pseudovector Vertices}
\smallskip

The pion and the pseudovector current do not couple to the scalar
diquark. Therefore, in both cases only those two contributions 
have to be computed which are obtained from the middle and lower diagrams in
figure\ \ref{emresolve} with replacing the photon-quark vertex by 
the pion-quark vertex of eq.\ (\ref{vertpion}), and by the pseudovector-quark
vertex of eq.\ (\ref{vertpv}), respectively. For Dirac part of the vertex
describing the pion coupling to the axialvector diquark this yields,
\begin{eqnarray}
  \tilde \Gamma^{\rho\lambda}_{5,ax} &= -&2\frac{m_q^2}{f_\pi}  
      \epsilon^{\rho\lambda\mu\nu} (p_d+k_d)^\mu Q^\nu \; \times 
             \label{pionaxres} \\
                  & &       \fourint{q} \frac{g_a^2\;P(q-Q/4)P(q+Q/4)}
             {(p_1^2+m_q^2)(p_2^2+m_q^2)(p_3^2+m_q^2)}  \; , \nonumber
\end{eqnarray}
and fixes its strength (at $Q^2=0$) to $\kappa_{ax}^5 \approx 4.5$, see
table~\ref{cc_2}. 

For the effective pseudovector-axialvector diquark vertex in eq.\
(\ref{vert5muax}) it is sufficient to consider the regular part, since its
pion pole contribution is fully determined by eq.~(\ref{pionaxres}) already.
The regular part reads,
\begin{eqnarray}
  \tilde \Gamma^{\mu\rho\lambda}_{5,ax}& =& 
            \fourint{q} \frac{g_a^2\;P(q-Q/4)P(q+Q/4)}
                        {(p_1^2+m_q^2)(p_2^2+m_q^2)(p_3^2+m_q^2)} \;\times \\
        & & \left[ -4 m_q^2 \epsilon^{\mu\rho\lambda\nu}\;(p_1+p_2+p_3)^\nu
               - \text{Tr}\gamma_5 \gamma^\rho \pslash_2 \gamma^\mu
                          \pslash_1 \gamma^\lambda \pslash_3 \right]. \nonumber
\end{eqnarray}
Although after the $q$-integration, the terms in brackets
yield the four independent Lorentz structures discussed in the paragraph
above  eq.\ (\ref{vert5muax}), only the first term contributes to 
$g_A(0)$ (with $p_1+p_2+p_3 = (3/4) (p_d+k_d) + q$).

The scalar-axialvector transition induced by the pion is described
by the vertex
\begin{eqnarray}
  \tilde \Gamma^{\rho}_{5,sa} & = & 4i\frac{m_q}{f_\pi}
   \fourint{q} g_s g_a\;P(q-Q/4)P(q+Q/4)\;\times \\
     & & \qquad \frac{(p_2p_3) p_1^\rho - (p_3p_1) p_2^\rho + 
               (p_1p_2) p_3^\rho}
            {(p_1^2+m_q^2)(p_2^2+m_q^2)(p_3^2+m_q^2)},
         \nonumber
      \label{pionsares}
\end{eqnarray}
and the reverse (axialvector-scalar) transition is obtained by
substituting $Q \rightarrow -Q$ (or $p_1 \leftrightarrow p_2$) in
(\ref{pionsares}). The corresponding vertex for the pseudovector
current reads
\begin{eqnarray}
  \tilde \Gamma^{\mu\rho}_{5,sa} &= -4im_q&
     \fourint{q} \frac{g_s g_a\;P(q-Q/4)P(q+Q/4)}
     {(p_1^2+m_q^2)(p_2^2+m_q^2)(p_3^2+m_q^2)} \times \nonumber \\
     & &
    \left[ \delta^{\mu\rho}( m_q^2 - p_1p_2 -p_2p_3-p_3p_1)\; 
     + \right. \label{pvsares} \\
    & & \left. \; \{p_1 p_2\}_+^{\mu\rho} + \{p_1 p_3\}_+^{\mu\rho}
     - \{p_2 p_3\}_-^{\mu\rho} \right]. \nonumber
\end{eqnarray}
The short-hand notation for a(n) (anti)symmetric product 
used herein is defined as $ \{p_1 p_2\}_\pm^{\mu\nu}=p_1^\mu p_2^\nu \pm 
p_1^\nu p_2^\mu$. The reverse transition is obtained
from $Q \rightarrow -Q$ together with an overall sign change 
in (\ref{pvsares}). As already mentioned in the main text,
the term proportional to $\delta^{\mu\nu}$  provides  99 per cent
of the value for $g_A$ as obtained with the full vertex. It
therefore clearly represents the dominant tensor structure.

\begin{table}
 \begin{center}
  \begin{tabular}{l|llll}
    &  $\kappa^5_{ax}$ & $\kappa^5_{\mu,ax}$ & $\kappa^5_{sa}$ 
    & $\kappa^5_{\mu,sa}$  \\ \hline
    & & & &  \\
 Set  I  & 4.53 & 4.41 & 3.97 & 1.97  \\
 Set  II & 4.55 & 4.47 & 3.84 & 2.13 
  \end{tabular}\\[+4pt]
  \caption{
          Strengths for 
          pion-- and pseudovector-diquark couplings.} 
  \label{cc_2}
 \end{center}
 \vspace{-0.2cm}
\end{table}
As explained for the electromagnetic couplings of diquarks, we use these
resolved vertices in connection with the rescaled couplings $g_s^{resc}$ and
$g_a^{resc}$ to compute $g_{\pi NN}$ and $g_A$ in the limit $Q\rightarrow 0$.
In this way the otherwise unknown constants that occur in the 
(pointlike) vertices of eqs.\ (\ref{vert5ax}--\ref{vert5musa}) are determined. 

As seen from the results in table\ \ref{cc_2},
the values obtained for these effective coupling constants are only slightly 
dependent on the parameter set (the only exception being
$\kappa^5_{\mu,sa}$ where the two values differ by 8\%). 
For the form factor calculations presented in Sec.~\ref{formfac_gp} we employ 
$\kappa^5_{ax}=4.5$, $\kappa^5_{\mu,ax}=4.4$, $\kappa^5_{sa}=3.9$ and
$\kappa^5_{\mu,sa}=2.1$.

\end{appendix}

\end{document}